\documentclass[letterpaper,titlepage,11pt]{article}

\pdfoutput=1

\usepackage{amsmath,amssymb,amsthm,mathrsfs,bbm}
\usepackage{latexsym,amscd,amsbsy,amsfonts,dsfont}
\usepackage{graphicx}
\usepackage{xcolor}
\usepackage[
      colorlinks=true,
      linkcolor=blue,
      urlcolor=blue,
      filecolor=black,
      citecolor=red,
      pdfstartview=FitV,
      pdftitle={},
        pdfauthor={},
        pdfsubject={},
        pdfkeywords={},
        pdfpagemode=None,
        bookmarksopen=true
      ]{hyperref}

\usepackage[utf8]{inputenc}
\usepackage{multirow}
\usepackage{makecell}
\usepackage{braket}
\usepackage{subcaption}
\usepackage{comment}
\usepackage{cite,mcite}

\newcommand{\be}{\begin{equation}}
\newcommand{\ee}{\end{equation}}
\newcommand{\beqa}{\begin{eqnarray}}
\newcommand{\eeqa}{\end{eqnarray}}
\newcommand{\bea}{\begin{eqnarray}}
\newcommand{\eea}{\end{eqnarray}}

\newcommand{\lp}{\left(}
\newcommand{\rp}{\right)}

\newcommand{\ord}[1]{{\mathcal O}\lp #1\rp}

\newcommand{\gsim}{\mathrel{\raisebox{-.6ex}{$\stackrel{\textstyle>}{\sim}$}}}

\def\clock{{\count0=\time
           \divide\count0 60
           \ifnum\count0<10 0\fi\the\count0
           \multiply\count0 -60 \advance\count0 \time
           :\ifnum\count0<10 0\fi \the\count0
         }}
\newcommand{\timestamp}{{\small\vbox{\hbox{\tt\jobname.tex}
\hbox{\the\day/\the\month/\the\year, \clock}}}}

\numberwithin{equation}{section}

\begin{document}

\begin{titlepage}
\leftline{}

\hfill	IFT-UAM/CSIC-21-144

\hfill	CPHT-RR107.122021
\vskip 2cm
\centerline{\LARGE \bf Holographic Complexity of Quantum Black Holes}

\bigskip

\vskip .9cm

\centerline{{\bf Roberto Emparan}$^{a,b}$, {\bf Antonia Micol Frassino}$^{b}$, {\bf Martin Sasieta}$^{c}$, {\bf Marija Toma\v{s}evi\'c}$^{b,d}$
}


\vskip 0.8cm
\centerline{\sl $^{a}$ Instituci\'o Catalana de Recerca i Estudis
	Avan\c cats (ICREA)}
\centerline{\sl Passeig Llu\'{\i}s Companys 23, E-08010 Barcelona, Spain}
\vskip 0.2cm

\centerline{\sl $^{b}$ Departament de F{\'\i}sica Qu\`antica i Astrof\'{\i}sica, Institut de
	Ci\`encies del Cosmos,}
\centerline{\sl  Universitat de
	Barcelona, Mart\'{\i} i Franqu\`es 1, E-08028 Barcelona, Spain}
\vskip 0.2cm

\centerline{\sl $^{c}$ Instituto de Física Teórica, IFT-UAM/CSIC,}
\centerline{\sl  Universidad Autónoma de Madrid, Nicolás Cabrera 13, 28049, Madrid, Spain}
\vskip 0.2cm
		
\centerline{\sl $^{d}$ CPHT, CNRS, Ecole Polytechnique, IP Paris, F-91128 Palaiseau, France}
\smallskip

\vskip 0.4cm
\centerline{\small\tt emparan@ub.edu, antoniam.frassino@icc.ub.edu,}
\vskip 0.2cm
\centerline{\small\tt martin.sasieta@csic.es, mtomasevic@icc.ub.edu}

\vskip 1.cm
\centerline{\bf Abstract} \vskip 0.2cm \noindent
We analyze different holographic complexity proposals for black holes that include corrections from bulk quantum fields. The specific setup is the quantum BTZ black hole, which encompasses in an exact manner the effects of conformal fields with large central charge in the presence of the black hole, including the backreaction corrections to the BTZ metric. 
Our results show that Volume Complexity admits a consistent quantum expansion and correctly reproduces known limits. On the other hand, the generalized Action Complexity 
picks up large contributions from the singularity, which is modified due to quantum backreaction, with the result that Action Complexity  does not reproduce the expected classical limit. Furthermore, we show that the doubly-holographic setup allows computing the complexity coming purely from quantum fields -- a notion that has proven evasive in usual holographic setups. We find that in holographic induced-gravity scenarios the complexity of quantum fields in a black hole background vanishes to leading order in the gravitational strength of CFT effects.

\noindent

\end{titlepage}
\pagestyle{empty}

\addtocontents{toc}{\protect\setcounter{tocdepth}{2}}

\tableofcontents
\normalsize
\newpage
\pagestyle{plain}
\setcounter{page}{1}

\section{Introduction}

	Different notions in many-body quantum chaos have gained prominence in AdS/CFT in the attempt to understand black holes at a fundamental level
	\cite{Sekino:2008he,Shenker:2013pqa,Maldacena:2015waa, Cotler:2016fpe}. 
	Among these ideas, perhaps the most intriguing one is the conjectural connection between quantum computational complexity and the emergence of the black hole interior geometry \cite{Susskind:2014rva,Stanford:2014jda,Roberts:2014isa,Susskind:2014jwa,Susskind:2014moa,Brown:2015bva,Brown:2015lvg, Brown:2017jil,Susskind:2018pmk}.
	However, the connection remains an incomplete entry of the holographic dictionary since the precise microscopic characterization of `holographic complexity' remains elusive. Current proposals identify a classical bulk observable that: (1) probes the black hole interior, and (2) has the characteristic phenomenology of a computational complexity.  Then, the value that such an observable takes in a bulk geometry defines the holographic complexity of the dual CFT state\footnote{
	See also \cite{Iliesiu:2021ari} for volume and spectral complexity.}.
	
	In this paper, we will probe in a new direction the two original proposals for the holographic complexity at a given time. The first one is the Volume Complexity (VC) of the state $\ket{\Psi}$, which is defined in terms of the volume of the extremal hypersurface $\Sigma$ anchored to the boundary timeslice,
	\be
	\mathcal{C}_{V}\left(\ket{\Psi}\right)\,= \;\dfrac{\text{Vol}(\Sigma)}{G\hbar \,L}\,,\hfill\label{VC}
	\ee
	where $L$ is a CFT length-scale, chosen to be parametrically $L \sim \ell_{\text{AdS}}$ (see also \cite{Couch:2018phr}). 
	The second one is the Action Complexity (AC) of the state $\ket{\Psi}$, which is  now covariantly characterized by the on-shell gravitational action $I$ evaluated on the Wheeler-deWitt (WdW) patch, $\mathcal{W}$, of the bulk spacetime for the given boundary time slice,
	\be
	\mathcal{C}_A\left(\ket{\Psi}\right)\,= \;\dfrac{I(\mathcal{W})}{\pi\hbar} \,.\hfill \label{AC}
	\ee
	
	Heuristically, VC is easier to grasp as it has the character of a computational complexity, specifically for a tensor network circuit that efficiently represents the entanglement structure of $\ket{\Psi}$ (see \cite{Swingle:2009bg,Almheiri:2014lwa,Pastawski:2015qua,Dong:2016eik, Bao:2018pvs} and references therein). For high-temperature thermofield double states, the presumed circuit is the Einstein-Rosen bridge, which, at late times \cite{Hartman:2013qma}, grows a number $ S = {\text{Area}}/{4G\hbar}$ of gates per thermal time $\beta = T^{-1}$, in accord with the general expectation about the complexity in these states. This linear growth is less trivial to extract from AC, which receives a non-vanishing contribution from the black hole singularity. Nevertheless, it follows the same law
	\be\label{latetimec}
	\left.\dfrac{d\,\mathcal{C}_{V,A}}{dt}\,\right |_{t \gg \beta} \;\sim \, TS\,,
	\ee
	up to $\mathcal{O}(1)$ coefficients that depend on the spacetime dimension and other details of the black hole. This behavior is the key property that sustains VC and AC as plausible measures of the complexity for the chaotic time-evolution of the black hole.
	
	\subsection*{Quantum bulk effects}
	
	The discussion so far implicitly assumes that the bulk is classical, which describes the CFT to leading order in the limit of large $N$, or more properly, central charge $c\to\infty$. Beyond this leading order in the large-$c$ expansion, several aspects of holographic complexity have been investigated in two-dimensional dilaton gravity \cite{Yang:2018gdb,Lin:2019qwu, Schneiderbauer:2019anh,Schneiderbauer:2020isp, Iliesiu:2021ari}, but its more general defining properties still await further elucidation.
	
	In this article, we shall consider a next step in the generalizations of \eqref{VC} and \eqref{AC}, which includes the leading finite-$c$ effects in the CFT, corresponding to the leading quantum corrections in the effective theory of the bulk. These incorporate effects beyond those considered in previous studies. 
	
	For VC, the tensor network rationale suggests a generalization of the schematic form
	\be\label{genVC}
	\mathcal{C}_V\left(\ket{\Psi}\right)\,= \;\dfrac{\text{Vol}(\Sigma)}{G\hbar\,L} \,+\,\dfrac{\delta \text{Vol}(\Sigma)\,+\, \mathcal{V}(\Sigma)}{G\hbar\,L}\,+\,\mathcal{C}_V^{\text{bulk}}\left(\ket{\phi}\right)\;+\;\dots\,.
	\ee
	Each term beyond the leading classical volume has a different origin. The first correction, $\delta \text{Vol}(\Sigma)$, represents the modification of the volume of the extremal bulk hypersurface $\Sigma$ due to the leading semiclassical backreaction on the bulk geometry\footnote{These were also included in \cite{Schneiderbauer:2019anh}.}. The next term, $\mathcal{V}(\Sigma)$, embodies the effect on $\Sigma$ of possible higher-curvature terms in the gravitational action generated from quantum renormalization (which also renormalizes the bulk coupling $G$). Finally, $\mathcal{C}_V^{\text{bulk}}\left(\ket{\phi}\right)$ accounts for the contribution of most interest to us: the complexity of quantum fields in the bulk, or more precisely, the `bulk tensor network' representation of the state $\ket{\phi}$ within the code subspace, in a suitable definition of complexity. The dots generically indicate higher-order terms in the bulk coupling as well as possible higher-curvature terms induced from, e.g., stringy effects. 
	The expression \eqref{genVC} bears structural resemblance to the quantum-corrected holographic entanglement entropy \cite{Faulkner:2013ana}.
	
	The structure of quantum corrections to AC is less clear. We shall think of them more abstractly, as
	\be\label{genAC}
	\mathcal{C}_A\left(\ket{\Psi}\right)\, = \;\dfrac{I(\mathcal{W})}{\hbar} \,+\,\delta \mathcal{C}_A\left(\ket{\Psi}\right)\,+\,\dots\,.
	\ee
	where $\delta \mathcal{C}_A\left(\ket{\Psi}\right)$ represents the $\mathcal{O}(1)$ AC functional in the large-$c$ expansion.
	
	\subsection*{Bulk CFT from double holography}
	
	At this stage, \eqref{genVC} is still largely a formal expression, in light of the absence of specific candidates for the higher-curvature correction $\mathcal{V}(\Sigma)$ and for the bulk complexity $\mathcal{C}_{V}^{\text{bulk}}(\ket{\phi})$ for weakly coupled fields. A way to circumvent the complicated problem of their definition in standard AdS/CFT systems is to instead consider theories in which gravity couples consistently to holographic conformal fields with central charge $c\,_{\text{bulk}}$. In the limit of large number of bulk quantum degrees of freedom, $c\,_{\text{bulk}}\rightarrow \infty$, the `braneworld holography'	description of such systems allows for consistent definitions of $\mathcal{V}(\Sigma)$ and $\mathcal{C}_{V}^{\text{bulk}}(\ket{\phi})$, in terms of classical magnitudes of the higher dimensional bulk.

\begin{figure}[!h]
		\centering
		\includegraphics[width = \textwidth]{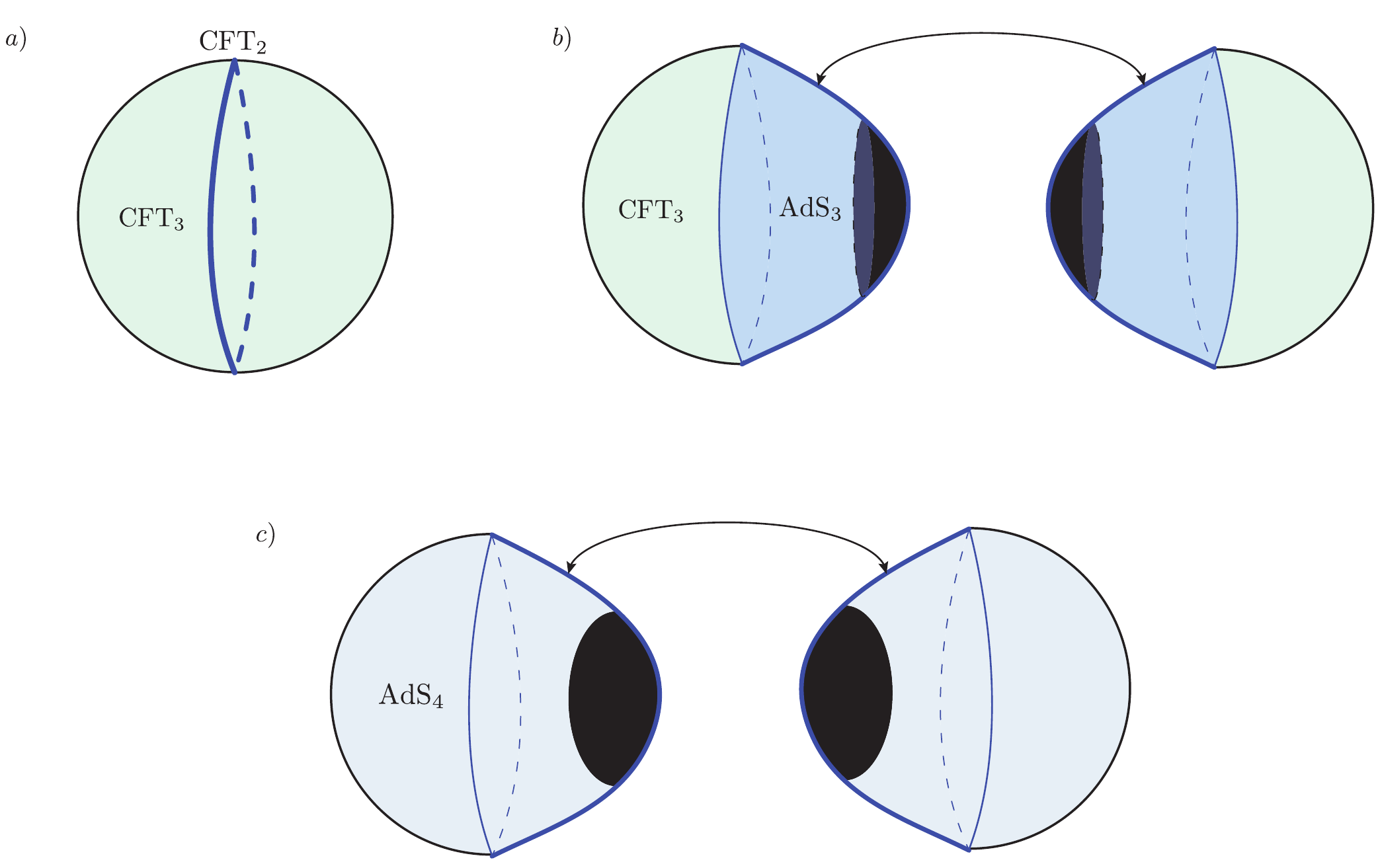}
		\caption{\small Sketch of a spatial slice in the holographic braneworld model under consideration. In $a)$ the UV description of the system is in terms of a thermal state of a dCFT$_3$. $b)$ In the brane perspective, the defect is replaced by gravity in AdS$_3$ with the CFT$_3$ as matter, which can leak towards the rigid CFT$_3$ via transparent boundary conditions. The left/right figures are identified along the AdS$_3$ region. The state of the system corresponds to a semiclassical black hole in thermal equilibrium, which consistently includes all of the planar quantum effects of the CFT$_3$ state. $c)$ In `double holography' the system is described in terms of AdS$_{4}$ gravity with a Planck brane. The state of the system consists of a four-dimensional bulk black hole which intersects the Planck brane. The left/right bulk patches are glued together along the brane via Israel's junction conditions.}
		\label{fig::defect}
	\end{figure}

	Holographic braneworld models implement these systems from a bottom-up perspective in a simple manner (see Fig.~\ref{fig::defect}). The idea is to consider the two possible dual descriptions of a defect CFT (dCFT) for a codimension-one conformal defect. The first such dual description replaces the defect by a certain gravitational theory in AdS with conformal matter which asymptotically couples to the rigid CFT. Alternatively, the doubly-holographic perspective is to view the previous effective description in terms of the localization of a higher-dimensional AdS gravity on a Planck brane \cite{Karch:2000ct,DeWolfe:2001pq}.
	
	Such models have proven to be useful for studying quantum black holes \cite{Emparan:2002px},  holographic entanglement entropy and its relation with black hole entropy \cite{Emparan:2006ni}, and more recently the island formula 
	\cite{Almheiri:2019hni,Rozali:2019day}. A higher-dimensional example in terms of the ground state of a dCFT was put to use in \cite{Chen:2020uac} as an avatar of the formation of entanglement islands in a simpler setup. Continuing in this direction, the so-called `subregion VC proposal' for the radiation \cite{Bhattacharya:2021jrn} has been also analyzed in these models, with particular focus on the analogue of the higher-curvature term $\mathcal{V}(\Sigma)$ in \eqref{genVC} for the ground state of a dCFT system \cite{Hernandez:2020nem}.
	
	In this paper, we employ double holography to define the VC of bulk quantum fields, $\mathcal{C}_V^{\text{bulk}}\left(\ket{\phi}\right)$. This allows a controlled calculation of it and of the backreacted volume $\delta \text{Vol}(\Sigma)$ in an explicit setup. In the same framework, we also investigate the properties of the AC complexity $\delta \mathcal{C}_A\left(\ket{\Psi}\right)$.
	
	Specifically, we consider a dCFT in three dimensions, where a CFT$_3$ with central charge $c_3$\footnote{This is a slight, but common, abuse of terminology, since there is no Weyl anomaly in three dimensions. The central charge is defined as a measure of the number of degrees of freedom, e.g., from the two-point function of the stress tensor or the free energy of the thermal state.} couples to the degrees of freedom in a two-dimensional defect described by a CFT$_2$ with central charge $c_2$. We take this system to be in a thermal state, which in the dual to the defect has a semiclassical black hole on the brane (see Fig. \ref{fig::defect}). Dualizing the entire dCFT to a four-dimensional bulk, the system can be studied using an exact solution for a black hole localized on a brane in AdS$_4$. The construction was initially described in \cite{Emparan:1999fd,Emparan:2002px}, and was revisited and extended in \cite{Emparan:2020znc}, where it was dubbed  \textit{quantum BTZ black hole} (quBTZ). The gravitational effects of the CFT$_3$ on the brane, with Newton's constant $G_3$ and AdS$_3$ radius $\ell_3$, are controlled by the parameter
	\begin{equation}\label{geff}
	    g_{\text{eff}}\sim \frac{G_3 c_3}{\ell_3}\sim\frac{c_3}{c_2}\,,
	\end{equation}
	which is required to be small for the consistency of the three-dimensional effective description.
	
	The specific choice of brane in our bulk construction has physical consequences. A minor one is that we take it to be two-sided, so the CFT$_3$ lives in a spatial 2-sphere and the defect is at an equatorial circle of it. In our solutions the two sides of the brane are identical copies, so we could just as well consider a one-sided end-of-the-world brane, and thus a boundary CFT, and simply double the number of degrees of freedom of the system. More importantly, we take the brane to be purely tensional, that is, its action consists only of the worldvolume area multiplied by the brane tension. Viewed as a holographic braneworld, this means that the three-dimensional gravity on the brane is entirely induced by the integration of UV degrees of freedom of the CFT$_3$ (i.e., the four-dimensional bulk portion that is excluded by the brane). What this implies for the holographic dCFT is in general less clear, but we will see that it has very concrete consequences for $\mathcal{C}_V^{\text{bulk}}\left(\ket{\phi}\right)$ -- namely, it is zero. Scenarios where gravity is not of the  induced-type are also possible,\footnote{In these scenarios, the brane action incorporates an intrinsic Einstein-Hilbert term \textit{à la} DGP \cite{Dvali:2000hr} which provides a bare gravitational coupling to the effective theory on the brane. This term modifies the RT prescription with an intrinsic area term on the brane \cite{Almheiri:2019hni,Chen:2020uac}. A similar effect is expected for holographic complexity \cite{Hernandez:2020nem}.} but on this matter, we will limit ourselves to a few comments near the end.

	In the case of completely induced gravity on the brane, the RT prescription in the classical four-dimensional bulk yields the entire entropy of the system from the area of the four-dimensional black hole, $S_{\text{gen}} = \text{Area}_4/4G_4\hbar$. In the three-dimensional theory, this quantity has the interpretation of a generalized entropy
	\be\label{sgen}
	S_{\text{gen}} = \,\dfrac{\text{Area}_3}{4G_3\hbar}\,+\,S_{\text{out}}\,+\,\dots\,,
	\ee
	where the dots represent higher-curvature Wald correction terms in the effective gravitational theory on the brane, and $S_{\text{out}}$ is the von Neumann entropy of the semiclassical state of the CFT$_3$ outside the horizon (including the bath). It was argued in  \cite{Emparan:2020znc} that this generalized entropy satisfies the first and second law of thermodynamics on the brane to all orders in the effective coupling $g_{\text{eff}}$ (in the large $c_3$ limit). Therefore, it is appropriate to regard it as measuring the number of degrees of freedom, or for our purposes, the effective number of links in the tensor network.
	
	In the same spirit, the holographic complexity of $\ket{\text{TFD}_\beta}$ is prescribed by \eqref{VC} and \eqref{AC} in the four-dimensional bulk, and these incorporate all quantum corrections of the CFT$_3$ on the brane in the limit $c_3\rightarrow \infty$.

	\subsection*{Generalized VC and AC}
	
	Using this framework, we will explicitly compute the generalized VC and AC, and test them for specific properties.
	Our first check will be to see whether the complexity is dominated by the classical bulk term when the bulk quantum effects are small. Although seemingly trivial, this will turn out to be a surprisingly subtle issue for AC. Another elementary test is whether the generalized complexities reproduce the expected late-time behavior
	\be
	\label{gengrowth} \left.\dfrac{d\,\mathcal{C}_{V,A}}{dt}\,\right |_{t \gg \beta} \;\overset{\textbf{?}}{\sim} \, TS_{\text{gen}}\,,
	\ee
	where the entropy is now the generalized one, $S_{\text{gen}}$.

	Our main results can be summarized as follows:
	
	\begin{itemize}
		\item{VC admits the structure of the quantum-corrected version \eqref{genVC} in the three-dimensional effective theory, to leading order in the effective coupling~\eqref{geff},  $g_{\text{eff}} \sim c_3/c_2$, with the leading term being the complexity of the classical  BTZ black hole. It also satisfies \eqref{gengrowth} at late times.}
		
		\item{AC, as defined by the four-dimensional bulk action, does \textit{not} reduce in a semiclassical expansion in  $g_{\text{eff}}$ to the classical three-dimensional AC plus quantum corrections. We find cancellations between the bulk, boundary, and joint terms in the action that render the late-time growth independent of $g_{\text{eff}}$. As a result, its value for $g_{\text{eff}}\rightarrow 0$ does not reproduce \eqref{latetimec} for the BTZ black hole. Hence this generalized AC does never reduce to an effective three-dimensional AC. We trace the origin of this discontinuity to the fact that the WdW patch reaches the inner singularity, whose structure changes qualitatively due to quantum backreaction, and which gives rise to large quantum contributions to AC.}
		
	\end{itemize}
	 These results highlight the sensitivity of the quantum-corrected AC functional \eqref{genAC} to the large quantum effects near the singularity. This does not seem to merely be a consequence of the doubly-holographic brane construction.	The value of AC for the ground state of a dCFT system has been observed in \cite{Chapman:2018bqj} to be insensitive to presence of the brane. Our results provide another piece of evidence of this phenomenon in a thermal state of a dCFT, but more importantly, they also point to a more fundamental characteristic of AC: it is a magnitude that allows to probe the black hole interior in a more precise way than VC.
		
	In addition to these results for the quBTZ solution, we are also able to test the generalized VC and AC in a class of three-dimensional black hole solutions, parametrically smaller than the AdS$_3$ radius, whose horizons are entirely due to quantum backreaction, and which in the absence of it become conical defects. When we study the AC and VC for these `quantum-dressed' conical defects, their complexity remains constant in the classical limit, since then the geometry is globally static. Therefore the complexity growth is entirely due to the quantum fields, although not directly from their bulk complexity $\mathcal{C}_V^{\text{bulk}}\left(\ket{\phi}\right)$ (which vanishes), but through their 
	backreaction on the geometry, which gives $\delta \text{Vol}(\Sigma)\neq 0$.

	\smallskip
	
	The remainder of the paper is organized as follows: In section \ref{sec::quBTZ}, we give a self-contained description of the static quBTZ black hole and its 4D bulk dual. In section \ref{sec::VC}, we analyze the VC of the dCFT system and identify the leading quantum-correction from the three-dimensional effective theory on the brane. In section \ref{sec::AC}, we analyze the late-time regime of AC, and find a discontinuity in its classical limit. We conclude in section \ref{sec:discuss}. The two appendices contain technical details.

	\section{Quantum BTZ black hole}
	\label{sec::quBTZ}
	
	We present here the main features of the quantum-corrected BTZ black hole (quBTZ), summarizing the analysis in \cite{Emparan:2002px,Emparan:2020znc}. 
	
	\subsection{Metric and parameters}
	
	The construction of the bulk geometry is based on the AdS$_4$ C-metric\footnote{AC in a different class of braneless (AdS) C-metrics has been considered in \cite{jiang2021holographic, Chen:2021qbs}. It is also possible to include a second brane at $r=\infty$ \cite{Emparan:1999fd}.}, which in \cite{Emparan:2020znc} was written in the form
	\be\label{cmetric}
	ds^2 \, = \, \dfrac{\ell^2}{(\ell + rx)^2}\left(-H(r)\,dt^2 \,+\,\dfrac{dr^2}{H(r)} \,+\,r^2\left(\dfrac{dx^2}{G(x)}\,+\,G(x)\,d\phi^2\right)\right)\,,
	\ee
	where
	\begin{gather}
	H(r) \, = \, \dfrac{r^2}{\ell_3^2}+\kappa-\dfrac{\mu\ell}{r}\,, \label{H}\\
	G(x) \, = \, 1-\kappa x^2-\mu x^3\,,
	\label{G}
	\end{gather}
	with $\kappa=\pm 1$.\footnote{$\kappa$=0 is also possible, but the results can be recovered as a limit of the other two cases.} This is an Einstein-AdS$_4$ metric with AdS radius
	\be\label{ell4}
	\ell_4 = \left( \frac{1}{\ell^2} + \frac{1}{\ell^2_3} \right)^{-1/2}\,.
	\ee
	The meaning of the parameters $\ell_3$, $\ell$ and $\mu$ will be clarified presently.
	
	To understand this geometry, note first that when $\mu=0$ the Riemann curvature is constant, so the spacetime is AdS$_4$ in particular coordinates. Let us consider the section $x=0$. It is immediately apparent that the three-dimensional geometry that is induced on it,
	\begin{equation}\label{btzcone}
	ds^2 \, = \, -\left(\frac{r^2}{\ell_3^2}+\kappa\right)\,dt^2 \,+\,\dfrac{dr^2}{\frac{r^2}{\ell_3^2}+\kappa} \,+\,r^2\,d\phi^2\,,
	\end{equation}
	is that of a BTZ black hole (when $\kappa=-1$) or a conical defect in AdS$_3$ (when $\kappa=+1$), with constant curvature radius $\ell_3$. The mass of the black hole and the conical defect depend on the periodicity of the angle $\phi$, which for now is arbitrary.  Since we will be mainly interested in BTZ-type black holes, in the following we set $\kappa=-1$, but later we will revisit the solutions with $\kappa=+1$.
	
	The section $x=0$ is where we will put a brane. To this effect, we keep the portion $x>0$ of the geometry and glue it at $x=0$ to an identical copy of the spacetime. The metric is continuous at the gluing surface, but the extrinsic curvature jumps when crossing it, which means that on this surface there is a brane with tension
	\begin{equation}
	\tau = \frac1{2\pi G_4\ell}\,.
	\end{equation}
	This gives the bulk interpretation of $\ell$. For our purposes, the primary theory is the one on the 3D brane, and the 4D bulk emerges holographically from it. Thus, we will often consider that $\ell_3$ remains fixed, and when we vary $\ell$ the bulk radius $\ell_4$ will change accordingly. Note that when $\ell$ is small, we have $\ell_4\simeq \ell$.
	
	Restricting to $x>0$, the range of $r$ is $(-\infty, -\ell/x)\cup(0,\infty)$.  The surface $r=\pm\infty$ in the bulk has AdS$_3$ geometry, and can be approached both from the side that contains the brane, where $r>0$, and from the side with $r<0$, which extends to the asymptotic boundary of the bulk geometry at $rx=-\ell$.
	
	\subsection{Quantum backreaction}
	
	The previous construction, with $\mu=0$, gives us a BTZ black hole on the brane, which extends into the bulk as a `BTZ black string'. However, our real interest is in the solutions where we turn on $\mu>0$, since these will have finite size black holes localized on the brane. We put this brane again at $x=0$ with the same tension as before, but now the geometry that is induced on it is not BTZ, but instead
	\begin{equation}\label{qubtz}
	ds^2 \, = \, -\left(\frac{r^2}{\ell_3^2}-1-\frac{\mu\ell}{r}\right)\,dt^2 \,+\,\dfrac{dr^2}{\frac{r^2}{\ell_3^2}-1-\frac{\mu\ell}{r}} \,+\,r^2\,d\phi^2\,.
	\end{equation}
	This differs from BTZ in several important respects. First, its curvature is not constant, so the geometry is not simply a quotient of AdS$_3$. At large $r$, the curvature radius asymptotes to $\ell_3$, but at $r=0$ there is a curvature singularity. Of more relevance to us, since $R_{ab}\neq -(2/\ell_3^2) g_{ab}$, this is not a solution to pure AdS$_3$ gravity. Instead, its correct interpretation is as a solution of a theory of three-dimensional gravity, with higher curvature terms, coupled to a large number of quantum conformal fields, namely, the holographic CFT$_3$ dual to the four-dimensional bulk. The effective action for this theory, which is derived from the four-dimensional classical gravitational action with a brane, is  
	\begin{equation}\label{3Deffact}
	I=\frac1{16\pi G_3}\int d^3 x \sqrt{-h}\left( \frac2{\ell_3^2} + R+\ord{\ell^2}\right) + I_\text{CFT}\,,
	\end{equation}
	where $G_3=G_4/2\ell_4$. The $\ord{\ell^2}$ terms are higher-curvature corrections to the gravitational action. Their explicit form can be found in \cite{Emparan:2020znc}, but we will not need them. The three-dimensional effective theory is most sensible when the gravitational effective action \eqref{3Deffact} is dominated by the Einstein-AdS terms, and therefore when $\ell\ll \ell_3$. 
	
	In this article, we do not work in any specific instance of AdS/CFT duality where the field theory for $I_\text{CFT}$ is explicitly known. Instead, for us the CFT$_3$ is holographically defined by the 4D bulk. 
	Its central charge is then given by
	\begin{equation}\label{c3}
	c_3 \sim \frac{\ell_4^2}{G_4} \sim \frac{\ell}{G_3}\,.
	\end{equation}
	The first expression here is the conventional entry in the AdS$_4$/CFT$_3$ dictionary, and the second one follows from \eqref{ell4} assuming that $\ell\ll \ell_3$. The value of the CFT action is $|I_\text{CFT}|\sim c_3$, while the typical value of the gravitational part of the action in \eqref{3Deffact} is 
	\begin{equation}\label{Igrav}
	   | I_\text{grav}|\sim \frac{\ell_3}{G_3}\,.
	\end{equation}
	Then, if we introduce a dimensionless `effective coupling'
	\begin{equation}
	   g_\text{eff}\sim  \frac{|I_\text{CFT}|}{|I_\text{grav}|}\sim \frac{G_3 c_3}{\ell_3}
	\end{equation}
	(see \eqref{geff}), which measures the size of effects of the CFT$_3$ on the background geometry, we find that 
	when $\ell\ll \ell_3$,
	\begin{equation}\label{Iratio}
	    g_\text{eff}\sim \frac{\ell}{\ell_3}\ll 1\,.
	\end{equation}
	In this regime the gravitational backreaction of the quantum fields will be small. Incidentally, the effects of quantum gravity in three dimensions are $\sim L_\text{Planck}^{(3)}/\ell_3=G_3/\ell_3$ and therefore negligible when $c_3\gg 1$.
	
	The AdS$_4$ C-metric with a brane gives a solution to this three-dimensional effective theory that captures (to all orders in $g_\text{eff}$) the backreaction of the quantum CFT$_3$ on a BTZ black hole, with metric \eqref{qubtz}. The stress tensor for the CFT$_3$ in this state is
	\begin{equation}\label{renst}
	\langle T^a{}_b \rangle\propto c_3\,\frac{\mu}{r^3}\,\text{diag}\{1,1,-2\}\lp 1+\ord{ g_\text{eff}}^2\rp\,.
	\end{equation}
	This indeed has the generic structure of the renormalized stress tensor of conformal fields in the presence of the BTZ black hole \cite{Steif:1993zv}. The stress tensor is naturally proportional to the central charge $c_3$, and its effect on the BTZ geometry will be $\propto c_3\, G_3\sim \ell$, which is small compared to the black hole horizon size $\sim \ell_3$.
	
	It will be important for us that the leading CFT effects come at linear order in $g_\text{eff}\propto \ell$, and thus can be disentangled from the higher-curvature corrections, which enter at $\ord{\ell^2}$. 
	
	\subsection{Properties of the solution}
	
	The parameter $\mu$, which from \eqref{renst} is seen to characterize the state of the quantum fields, depends on the mass $M$ of the black hole. In \eqref{btzcone} and \eqref{qubtz}, the mass parameter is not apparent, since these metrics are not written in the canonical form where the periodicity of $\phi$ would be $2\pi$. Instead, in the coordinates of \eqref{btzcone} and \eqref{qubtz} the mass varies with the periodicity of the identification $\phi\sim\phi+2\pi\Delta$ as\footnote{This relation holds when the gravitational theory is Einstein-AdS$_3$. The higher-curvature corrections in the effective theory on the brane introduce a modification of the mass formula that,  when $\ell$ is small, is $\ord{\ell^2}$ \cite{Emparan:2020znc}.}
	\begin{equation}\label{MDelta}
	M=\frac{\Delta^2}{8G_3}\,.
	\end{equation}
	Now, when $\mu>0$ the periodicity $\Delta$ of the solution cannot be chosen arbitrarily; conversely, the state of the quantum CFT$_3$, which is parametrized by $\mu$, is not arbitrary but determined by the geometry of the black hole with mass $M$. In the dual bulk description of the CFT$_3$, this comes about as a consequence of bulk regularity. The four-dimensional geometry describes a black hole that is stuck to the brane, in the shape of a hemi-spherical cap or finite cigar, instead of extending indefinitely away from the brane as a black string (see Fig.~\ref{fig::defect}). Regularity at the axis of rotational symmetry in the bulk (where $G(x)=0$) requires that we fix $\Delta$ to be
	\begin{equation}\label{Deltax1}
	\Delta=\frac2{G'(x_1)}=\frac{2 x_1}{3+x_1^2}\,,
	\end{equation}
	where $x_1$ is the smallest positive root of $G(x)$, so that
	\begin{equation}\label{mux1}
	\mu=\frac{1+x_1^2}{x_1^3}\,.
	\end{equation}
	These two equations determine a relation (parametrized by $x_1$) between $\Delta$ and $\mu$, and hence between the black hole mass $M$ and the quantum state parameter $\mu$. 
	
	It turns out that for each value of $M\in [0,1/(24G_3)]$ there are two possible values of $\mu$, while for $M>1/(24 G_3)$ there are none. That is, in the lower mass range there are two possible configurations of black holes with the same mass but different state of the CFT$_3$, and (since the CFT$_3$ backreacts differently on them) different entropy. In \cite{Emparan:2002px,Emparan:2020znc}, these were interpreted as corresponding to states where the CFT$_3$ is dominated by large Casimir effects in one case, and by thermal (Hawking) effects in the other, with the former having larger entropy. Instead, for black holes with masses  $M>1/(24 G_3)$ the conformal fields are not excited, which is a peculiarity of the holographic theory.
	
	\subsection{Generalized entropy}
	
	As we mentioned in the introduction, according to the holographic interpretation, the area of the four-dimensional bulk horizon gives the total (or generalized) entropy of the system, which consists of the Bekenstein-Hawking-Wald entropy of the three-dimensional black hole \eqref{qubtz} and the entropy of the quantum CFT$_3$, \eqref{sgen}. Expanding in $\ell$, the structure of this entropy is
	\begin{equation}
	S_{\text{gen}} =\frac{\text{Area}_3}{4G_3}+\ell\frac{\delta \text{Area}_3}{4G_3}+\ell^2 \mathcal{S}+S_{\text{out}}(|\phi\rangle)+\dots\,,
	\end{equation}
	which is written emphasizing the similarity to the Volume Complexity in \eqref{genVC}: the first term is the area entropy of the uncorrected BTZ black hole; the second, the modifications to its area due to leading-order quantum backreaction; the third term accounts for the Wald entropy from higher-curvature terms in the action \eqref{3Deffact}; the last term is the quantum entropy from the entanglement of the conformal fields outside the horizon, which starts at linear order in $\ell$.
	
	For the quBTZ solution with mass $M$, expanding to linear order in $\ell$ one finds that
	\begin{equation}
	S_{\text{gen}} =\frac{\text{Area}_4}{4G_4}= \pi\ell_3\sqrt{\frac{2M}{G_3}}\left(1+\frac{\ell \mu}{2\ell_3}\right)
	-\frac{\pi\ell}{x_1}\sqrt{\frac{2M}{G_3}}+\ord{\ell^2}
	\end{equation}
	(the functions $x_1(M)$ and $\mu(M)$ are given implicitly by \eqref{MDelta}, \eqref{Deltax1}, and \eqref{mux1}).
	The leading order term is the Bekenstein-Hawking entropy of the BTZ black hole with mass $M$. The term $\propto \ell\mu$ inside the brackets is the correction to the area of the BTZ black hole due to quantum backreaction on the geometry. The last term then corresponds to the quantum entanglement entropy of the conformal fields.
	
	\subsection{Quantum dressing}
	
	Let us briefly mention the solutions with  $\kappa=+1$. The corresponding three-dimensional geometries on the brane have masses in the range $-1/(8G_3)<M<0$ where, in the classical theory, only conical defects exist. The AdS C-metric with $\mu>0$ holographically describes how these conical singularities get dressed with a black hole horizon through the quantum backreaction of conformal fields (whose stress tensor comes from the Casimir effect in a cone). Therefore, even if they cannot be interpreted as quantum-corrected BTZ black holes, they are nevertheless valid semiclassical black holes to which our analysis can be applied.
	
	Since the horizon in these solutions disappears in the classical limit $\ell\to 0$, they can be regarded as fully quantum black holes, which makes them appealing. Their main properties have been studied in \cite{Emparan:2002px,Emparan:2020znc}.

	\subsection{Defect CFT}
	
	Above we have emphasized the interpretation of the construction as yielding a solution for the quantum-corrected black hole on the three-dimensional brane at $x=0$. However, as we discussed in the introduction, in our setup this system is not closed, but coupled at its asymptotic boundary (at $r\to\infty$) to a `bath' CFT$_3$ in a non-dynamical background (the conformal geometry at $rx=-\ell$). The transparent conditions at the boundary allow the two subsystems to exchange energy, and the bath CFT$_3$ has both a thermal component $\sim \{-2,1,1\}$ and a Casimir component $\sim \{1,1,-2\}$ \cite{Hubeny:2009kz}, the latter dominating when $g_\text{eff}\ll 1$.
	
	When the three-dimensional gravitating brane is dualized to a two-dimensional CFT$_2$ residing at its boundary, the latter couples to the non-gravitating CFT$_3$ as a defect (see Fig.~\ref{fig::defect}a). The bulk black hole holographically puts this defect CFT in a thermal state.
	
	The standard AdS$_3$/CFT$_2$ dictionary assigns to the defect a central charge
	\begin{equation}
	    c_2=\frac{3\ell_3}{2G_3}\,.
	\end{equation}
	This is of course the characteristic value of the three-dimensional gravitational action, \eqref{Igrav}, so we see that the effective coupling of the CFT to gravity
	is
	\begin{equation}\label{geff2}
	    g_\text{eff}\sim \frac{\ell}{\ell_3}\sim \frac{c_3}{c_2}\,.
	\end{equation}
	Thus, in the regime of interest to us we have $c_3\ll c_2$.
	
	\medskip
	
	We now turn to evaluating different proposals for holographic complexity in this solution.
	
	\section{Volume Complexity}
	\label{sec::VC}

	We adopt the prescription that the VC of the time-evolved thermofield double state $\ket{\text{TFD}_\beta(t)}$ of the pair of dCFTs is given by the extremal volume of $\Sigma_t$ in the four-dimensional bulk
	\be\label{VC4}
	\mathcal{C}_V(t)\, = \;\dfrac{\text{Vol}(\Sigma_t)}{G_4 L}\,.
	\ee
	The slice $\Sigma_t$ is anchored to the left and right asymptotic boundaries at $t_L = t_R = t$. Here we are choosing the standard time evolution with the Hamiltonian $H = H_L+H_R$. We shall conveniently fix $L = 4\pi \ell_{3}$ as the characteristic length scale of the holographic dCFT system.
	
	If we were to evaluate the VC of the brane system using the standard prescription, we would have employed the volume of a section in the three-dimensional brane. The difference between \eqref{VC4} and the standard VC formula captures the quantum corrections to VC on the brane to leading order in the large $c_3$ limit. In this section, will show that \eqref{VC4} admits the expansion
	\be\label{VC3}
	\mathcal{C}_V(t)\, = \; \mathcal{C}^{\text{BTZ}}_V(t)\,+\, \mathcal{C}^{q}_V(t) +\ord{g_{\text{eff}}}^2\,. 
	\ee
	This functional of the three-dimensional effective theory has the structure of \eqref{genVC} where the dominant term $\mathcal{C}^{\text{BTZ}}_V$ corresponds to the VC of the classical BTZ black hole. This term is proportional to $c_2$. The next term scales with $c_3$ (so it is smaller than the previous one by $g_\text{eff}\sim c_3/c_2$) and corresponds to the leading quantum correction in the large-$c_3$ limit. This correction can be further expanded in powers of $g_{\text{eff}}$, and we shall just consider the leading term $\mathcal{C}^{q}_V$. As we have emphasized, in our setup this term receives contributions from the quantum backreaction, $\delta \text{Vol}(\Sigma_t)/G_3 \ell_3$, and from the complexity of the state of the bulk quantum fields $\mathcal{C}_V^{\text{bulk}}(\ket{\phi})$, but not from the higher-curvature terms of the 3D effective theory, which enter at order $g_{\text{eff}}^2$.

	\subsection{Extremal volume}
	
	Let us write the VC functional \eqref{VC4} for our specific solution \eqref{cmetric}. It will be convenient to define the coordinate 
	\begin{equation}
	    z = x\,r 
	\end{equation}
    and take $X^\mu = (t,r,z,\phi)$ as our coordinates. The brane then lies at $z=0$.
	We denote the embedding functions of a spacelike hypersurface $\Sigma$ in the four-dimensional bulk by $y^a = y^a(X^\mu)$, $a=1,2,3$. The tangent vectors are $e_a^\mu\, = \, \partial_a X^\mu$ and the induced metric on $\Sigma$ is $h_{ab}\,= \, g_{\mu\nu}\,e_a^\mu e_b^\nu$.

	From axial symmetry, we can take $y^3 = \phi$, and restrict the extremization problem to the family of axially-symmetric hypersurfaces, $\partial_\phi X^\mu = \delta_\phi^\mu$. We will moreover take $y^1 = r$ and $y^2 = z $ as the two remaining parameters, and define $\dot{X}^\mu = \partial_r X^\mu$ and $X^\mu \, '   = \partial_z X^\mu$. In these coordinates, the induced metric of the hypersurface reads
	\begin{gather}
	ds^2_{\Sigma}\, = \, \dfrac{\ell^2}{(\ell+z)^2}\,\left[\left(-H\,\dot{t}^2\,+\,H^{-1}\,+\,\dfrac{z^2}{Gr^2}\right)dr^2\,+\,\left(-H\,t'^2\,+\,G^{-1}\right)dz^2\,+\,\right.\nonumber\\[.4cm] \left.-2\left(H\,\dot{t}\,t'\,+\,\dfrac{z}{Gr}\right)\,dr\,dz\,+\,r^2G\,d\phi^2\right]\,.
	\end{gather}
	
	The VC of the system \eqref{VC4} is then obtained by finding the extremal value of the functional
	\begin{equation}\label{volfunct}
	\mathcal{C}_V(t)\, \, =  \;\text{ext} \,\left\lbrace\,\dfrac{\ell\Delta}{G_4\ell_3}\int dr\,dz\;\,\dfrac{\ell^2\,r}{(\ell+z)^3} \, \sqrt{-H\,\left(\dot{t}+\dfrac{zt'}{r}\right)^2\,+\,H^{-1}\,-\,G\,t'^2}\,\right\rbrace\,,
	\end{equation} 
	over the embedding function $t = t(r,z)$. We have added an extra factor of $2$ to restrict to reflection-symmetric hypersurfaces across the brane, and implicitly restricted the domain of integration to include only one side of the brane, $x >0$, so that even if $r$ and $z$ can be negative, we always have $z/r>0$.
	
	The boundary conditions of this extremization problem are by definition of Dirichlet type at $z = -\ell$ for each asymptotic boundary. Instead of considering these for the other side of the brane, we can equivalently impose Neumann boundary conditions on the brane. These arise from the extremality of the volume of $\Sigma_t$ under arbitrary reflection-symmetric deformations $\delta y^a(X^\mu)$ which are tangent to the brane at $z=0$. From local extremality of $\Sigma_t$, the effect of these deformations for $z>0$ vanishes, and the total volume variation reduces to a surface integral $\int dS_a\delta y^a$ on $\Sigma_t \cap \textbf{brane}$, with a surface element $dS_a$ tangent to $\Sigma_t$. For the volume to remain extremal, $dS_a$ must be orthogonal to an arbitrary tangent deformation $\delta y^a$ on the brane, i.e. it has to be orthogonal to the brane. Equivalently, the normal $N_{\Sigma_t} \propto (dt-\dot{t}\,dr-t'\,dz)$ is required to lie on the brane at $z=0$, which is achieved from the requirement
	\begin{equation}\label{bc}
	t'(r,0) = 0\,.
	\end{equation}
	At this point it may seem to be a mere technicality, but in Sec.~\ref{subsec:qucorr} we will find that this boundary condition, which follows from the purely tensional character of the brane, has important physical consequences.

	The extremization \eqref{volfunct} involves a complicated PDE problem that in general requires numerical resolution. Here, instead, we will compute the leading quantum correction $\mathcal{C}^{q}_V(t)$ analytically by expanding the VC functional to leading order in $g_{\text{eff}}$.

	\subsection{Classical term}
	
	We will now show that the classical VC of the BTZ black hole in \eqref{VC3} is recovered from  \eqref{volfunct} in the  limit $\ell \rightarrow 0 $, with $\ell_{3}$ and $G_3$ fixed, so that the coupling $g_{\text{eff}}$ vanishes. In this limit, the doubly-holographic bulk description disappears, since $G_4 \sim 2\ell G_3$ and $\ell_4 \sim \ell$ vanish. At the same time, the induced metric on the brane becomes a classical BTZ black hole. Observe that we are keeping $c_2$ finite while $c_3$ vanishes. A different limit where also $g_\text{eff}\sim c_3/c_2\to 0$ keeps instead $c_3$ finite with $c_2\to\infty$, and corresponds to having the BTZ black hole at the non-dynamical boundary of AdS$_4$, in which case we do not recover a finite complexity of BTZ.

	Naively, \eqref{VC4} and \eqref{volfunct} imply that the VC vanishes as $\ell^2$ in this limit. However, this is not the case due to the fact that in this limit the conformal factor of the C-metric in \eqref{volfunct} localizes the functional to the brane. The relevant term for this effect in \eqref{volfunct} is
	\be\label{conffact}
	g(\ell,z)\, = \, \dfrac{\ell^2}{(\ell+z)^3}\,.
	\ee
	
	The function $g(\ell,z)$ has a regular power series in $\ell$ (starting at $\ell^2$) for $|\ell| > |z|$, but this is not the appropriate regime in the limit that we want to consider here. In fact, as we show in Appendix \ref{appendix::A}, in our limit the function becomes a Dirac delta
	\be\label{delta}
	\lim\limits_{\ell\rightarrow 0}\;g(\ell,z)\, = \, \delta(z) \,,
	\ee
	in the distributional sense, for nice enough test functions.
	
	The conformal factor will therefore localize the volume functional \eqref{volfunct} in this limit to the brane. Commuting the limit $\ell \rightarrow 0$ with the extremization in \eqref{volfunct}, and using \eqref{delta}, we get
	\begin{equation}\label{volfunctzeroth}
	\lim_{\ell \rightarrow 0}\,\mathcal{C}_V(t) \, =  \;\underset{\sigma}{\text{ext}} \,\left\lbrace\,\dfrac{\Delta}{4G_3\ell_3} \,\int dr\,r\, \sqrt{-H_0\,\dot{t}^2\,+\,H_0^{-1}}\right\rbrace\,\;\,,
	\end{equation} 
	where $H_0(r) = r^2/\ell_{3}^2-1$ is the redshift factor of the BTZ black hole, and the extremization is performed over the intersection 
	\begin{equation} \label{Sigma-brane}
	\sigma = \Sigma \cap \textbf{brane}
	\end{equation}
	with the embedding function $t= t(r,0)$. This functional coincides precisely with the volume of an axially-symmetric spatial slice in the geometry of a BTZ black hole. The factor $\Delta$ is due to the integration over $\phi$. Moreover, the slices $\sigma$ are all anchored to the asymptotic boundary at $r=\infty$ on both sides, so \eqref{volfunctzeroth} is indeed the VC of the thermofield double of the two CFT$_2$'s (which are not defects anymore in this limit)
	\be
	\lim_{\ell \rightarrow 0}\,\mathcal{C}_V(t) \, =\,\left.\dfrac{\text{Vol}(\sigma_t)}{8\pi G_3\ell_3}\,\right|_{\ell =0} =\, \mathcal{C}^{\text{BTZ}}_V(t)\,,
	\ee
	where $\sigma_t$ is the extremal volume slice on the brane. Thus, in this limit, we recover the classical contribution to the VC of the system from the brane perspective \eqref{VC3}.

	\subsection{Leading quantum correction}
	\label{subsec:qucorr}

	Next, we show that the leading quantum correction $\mathcal{C}^{q}_V(t)$ in \eqref{VC3} can be split into two terms, each of which can be attributed to the two components of the CFT$_3$: at the non-gravitating boundary of AdS$_4$, and on the gravitating brane. More specifically, the first term is UV-divergent and corresponds to the short-range VC of the rigid CFT$_3$ bath. The second term makes up for the effect of the CFT$_3$ on the brane, and it is, in principle, a combined effect of the semiclassical backreaction of the CFT$_3$, $\delta \text{Vol}/G_3\ell_{3}$, and of the complexity of the bulk quantum fields, $\mathcal{C}_V^{\text{bulk}}(\ket{\phi})$. As we will see, the second term is the one 
	we are after.
	
	The short-range VC of the CFT$_3$ bath arises from the volume of $\Sigma_t$ close to the asymptotic boundary  of AdS$_4$. As we explained, the AdS C-metric continues beyond $r=\infty$ into negative values of the radial coordinate, and the asymptotic boundary lies in this domain. Isolating the $r<0$ part of the VC functional \eqref{volfunct}, this term reads
	\be\label{r<0}
	\mathcal{C}_{\text{UV}}\,=\,\left.\lim_{\varepsilon\rightarrow 0}\,\dfrac{2\Delta\ell}{G_4\ell_3} \,\int_{-\infty}^{-\frac{\ell+\varepsilon}{x_1}} dr\,\int^{x_1 r}_{-\ell+\varepsilon}\,dz\;\,\dfrac{\ell^2\,r}{(\ell+z)^3} \, \sqrt{-H\left(\dot{t}+\dfrac{zt'}{r}\right)^2\,+\,H^{-1}\,-\,G\,t'^2}\,\right|_{\Sigma_t}\,,
	\ee
	where we have introduced a bulk regulator near the boundary at $z = -\ell + \varepsilon$, and there is a factor of $2$ coming from the two asymptotic boundaries.  Note that this term does not contribute classically, since the localization of the conformal factor \eqref{delta} in the limit $\ell\to 0$ excludes a contribution from negative values of $z$.\footnote{The right order of limits is $\varepsilon \rightarrow 0$ before $\ell\rightarrow 0 $.}
	
	This contribution to the complexity, although divergent, is not of interest to us. Its structure when $\varepsilon \rightarrow 0$ follows the general analysis of \cite{Carmi:2016wjl}, and is understood as the complexity associated to the short-range correlations of the CFT$_3$ in the rigid bath and in the asymptotic region of the brane. This piece is the dominant part of the contribution from the region \eqref{r<0}, all of which is outside the black hole, and since this region is static, it simply adds a constant to the VC at late times. Therefore, we shall not consider it anymore.

	The most interesting term for us is the proper quantum correction to VC from the CFT$_3$ on the brane. This term is obtained by expanding the integrand of the VC functional \eqref{volfunct} to linear order in $\ell$, which necessarily means going beyond \eqref{delta} for the conformal factor. In Appendix~\ref{appendix::A}, we show that
	\be\label{deltanext}
	g(\ell,z)\, = \, \delta(z)\,-\,\ell\,\dfrac{d}{dz}\delta(z)\,+\,\mathcal{O}(\ell^2)\,,
	\ee
	in the distributional sense. 
	
	The new term linear in $\ell$ in \eqref{deltanext} also localizes the volume functional onto the brane. After integrating by parts, the integrand reads
	\be\label{normvanish}
	\delta(z)\,\dfrac{d}{dz} \left(\sqrt{-H\left(\dot{t}+\dfrac{zt'}{r}\right)^2\,+\,H^{-1}\,-\,G\,t'^2}\,\right) = \, 0\,,
	\ee
	which vanishes by virtue of the boundary condition \eqref{bc} and $G'(0) = 0$. We can view this effect geometrically by realizing that the family of hypersurfaces $\Sigma$ is reflection-symmetric across the brane, and the boundary condition \eqref{bc} moreover imposes one-time differentiability in the direction orthogonal to the brane. Since all the other properties of the induced volume refer to the tangent directions to the brane, its normal derivative on the brane vanishes for any $\Sigma$, which is what \eqref{normvanish} reflects. 
	
	As a result, the contribution linear in $\ell$ to the integrand of \eqref{volfunct} is obtained by taking the leading, $\ell^0$ term in \eqref{deltanext}, and extracting the linear backreaction correction from the CFT$_3$ on the geometry, specifically, the redshift factor $H(r)$ of the quBTZ black hole. We can then conclude that the leading order quantum correction $\mathcal{C}^{q}_V(t)$ is given by the two terms
	\be\label{qc}
	\mathcal{C}_V(t)\, =\,\mathcal{C}^{\text{BTZ}}_V(t)\,+ \,\mathcal{C}_{\text{UV}}\,+\,\dfrac{\delta \text{Vol}(\sigma_t)}{8\pi G_3 \ell_3 }\,+\,\ord{g_{\text{eff}}}^2\,,
	\ee
	where $\delta \text{Vol}(\sigma_t)$ is the $\mathcal{O}(\ell)$ change in volume of the extremal surface $\sigma_t$ on the brane, due to the semiclassical backreaction of the CFT$_3$. As we mentioned, the term $\mathcal{C}_{\text{UV}}$ in \eqref{r<0} does not add to the complexity of the state; instead, it pertains to the (static) UV degrees of freedom of the CFT$_3$ in the bath and in the brane far from the black hole, so we will ignore it. On the other hand, we have seen that to linear order in $\ell$, the infrared, time-dependent complexity of the dCFT is localized on the brane, so we can conclude that this complexity must be attributed to the defect. To quadratic order in $\ell$ this localization does not happen anymore, and therefore at that order the bath also contributes to time-dependent complexity.
	
	It may seem surprising that the contribution to the complexity from the bulk quantum fields vanishes in \eqref{qc}. However, this is not entirely accurate. The complexity of these quantum fields must be thought of as consisting of two parts: one of them arises from the CFT degrees of freedom above the cutoff, and is naturally absorbed in a renormalization of $G_3$, which is automatically incorporated in the doubly-holographic construction. The other term, which we refer to as $\mathcal{C}_V^{\text{bulk}}(\ket{\phi})$, is the complexity of the CFT degrees of freedom below the cutoff and after renormalization. It is the latter that vanishes for the particular state of the dCFT under consideration. 
	
	This feature of VC generalizes to $\mathbf{Z}_2$-symmetric states of holographic dCFT systems, as long as the brane is purely tensional.
	Besides the semiclassical backreaction on the geometry on the brane, the leading local corrections in $\ell/\ell_3$ to the extremal volume come from the normal derivative of $\Sigma_t$, which vanishes on such branes.
	
	This means that if the action of the brane incorporated an intrinsic Einstein-Hilbert term, then a modification of the standard VC prescription in the doubly-holographic bulk would be needed (see \cite{Hernandez:2020nem}). The corresponding hypersurface $\Sigma_t'$ would not be orthogonal to the brane, which we expect results in a non-vanishing quantum bulk complexity $\mathcal{C}_V^{\text{bulk}}(\ket{\phi})$ at leading order in $g_{\text{eff}}$.

	\subsection{Semiclassical backreaction on the volume}
		\label{subsec:volume qbtz}
	
	We now proceed to compute the leading order semiclassical backreaction to the extremal volume on the brane, ${\delta \text{Vol}(\sigma_t)}/{8\pi G_3 \ell_3 }$. By virtue of \eqref{qc}, this quantum effect encapsulates all of the quantum corrections to VC coming from the brane at leading order in $g_{\text{eff}}$. 
	
	We will consider the extremal volume slice $\overline{\sigma}_t$ in the quBTZ metric, i.e., to all orders in the backreaction of the CFT$_3$ on the brane, and then expand the results to the leading order backreaction, 
	\begin{equation}
	    \text{Vol}(\overline{\sigma}_t) = \text{Vol}(\sigma_t) + \delta \text{Vol}(\sigma_t) + \mathcal{O}(g_{\text{eff}})^2\,.
	\end{equation}
	We can restrict to circularly symmetric surfaces in the metric~\eqref{qubtz}, such that the volume of $\overline{\sigma}_t$ is obtained from the extremization of
	\be\label{onshellv}
	\text{Vol}(\overline{\sigma}_t) \, =\,2\pi\Delta\,\text{ext}\left\lbrace \,\int  d\lambda \,\mathcal{L}_V \right\rbrace\,,
	\ee
	with volume density
	\be
	\mathcal{L}_V = \,r\,\sqrt{-H(r)\,\dot{t}^2\,+\,\frac{\dot{r}^2}{H(r)}}\,,
	\ee
	over the embedding functions $t=t(\lambda),\; r=r(\lambda)$. Here $\lambda$ is a parameter that measures radial displacement along the surface, and $\dot{t} = dt/d\lambda$, $\dot{r} = dr/d\lambda$. The boundary conditions are of Dirichlet type, $r(\lambda^{L,R}_\infty) = r_\infty$ and $t(\lambda^{L,R}_\infty) = t$, where the parameter $\lambda \in (\lambda^{L}_\infty,\lambda^{R}_\infty)$ runs from the left to the right asymptotic boundary.
	
	Since we are considering static configurations, the canonical conjugate momentum $P_t$ will be conserved along the extremal $\overline{\sigma}_t$. Choosing a parametrization $\lambda$ such that the volume density is constant, $\mathcal{L}_V = 1$, we have
	\begin{equation}\label{momentum}
	P_t \equiv \frac{\partial \mathcal{L}_V}{\partial \dot{t}} = {-r^2\,\dot{t}\, H(r) }\,.
	\end{equation}

	This conserved quantity, together with the choice $\mathcal{L}_V = 1$, allows to write down first order equations for the embedding function $r(\lambda)$, which read
	\begin{equation}\label{eom}
	\dot{r}^2 = \frac{H(r)}{r^2} + \frac{P_t^2}{r^4}\,,
	\end{equation}
	and the conserved momentum $P_t$ can be evaluated on the section where $\overline{\sigma}_t$ attains its minimum radius $r_{\text{min}}$,
	\be\label{momentummin}
    P_t \, = \, r_{\text{min}}\sqrt{-H(r_{\text{min}})}\,.
	\ee
	
	The asymptotic anchoring time $t$ is implicitly related to the minimum radius $r_{\text{min}}$ through the boundary condition $t(\lambda^{L,R}_\infty) =t$. Integrating over $\dot{t}$ in \eqref{momentum}, this can be expressed as
	\be\label{timequBTZ}
	t\, = \, -\int_{r_{\text{min}}}^{r_\infty}\dfrac{P_t\,dr}{H(r)\sqrt{r^2H(r)+P_t^2}}\,.
	\ee
	Similarly, using the integrated version of \eqref{eom}, we can write the on-shell volume \eqref{onshellv} as the radial integral 
	\be\label{onshellv2}
	\text{Vol}(\overline{\sigma}_t) \, =\,2\pi\Delta \,\int_{\lambda_\infty^{L}}^{\lambda_\infty^{R}}  d\lambda \, = \,4\pi\Delta \int_{r_{\text{min}}}^{r_\infty}\dfrac{r^2dr}{\sqrt{r^2H(r)+P_t^2}}\,,
	\ee
	where we introduced a bulk regulator $r_\infty$ close to the asymptotic boundary, and added a factor of $2$ to incorporate the left/right reflection symmetry of $\overline{\sigma}_t$. 
	
	The bulk regulator $r_\infty$ is fixed since backreaction effects decay sufficiently fast as $r\rightarrow \infty$ \cite{Emparan:2021yon}. We follow the procedure in \cite{Chapman:2016hwi} and use the Fefferman-Graham expansion close to the asymptotic boundary, in which the metric must look like 
    \be\label{FGBTZ}
    ds^2\, \approx \, \ell_3^2\,\dfrac{dz^2+\ell_3^2\,d\overline{\phi}^2}{z^2}\,,
    \ee
    where $z$ is the Fefferman-Graham radial coordinate and $\overline{\phi}\sim \overline{\phi}+2\pi$. Comparing \eqref{FGBTZ} with the asymptotics of the quBTZ metric, we identify $\phi = \overline{\phi} \Delta$ and $r = \ell_3^2/(z\Delta)$. The asymptotic cutoff is defined at $z = \varepsilon$, which in our coordinates is then $r_\infty = \ell_3^2/(\varepsilon{\Delta})$.
	
	In order to solve \eqref{onshellv2} for $\delta \text{Vol}(\sigma_t)$, we will expand the quantities $r_{\text{min}} = r_{\text{min}}^0+\delta r_{\text{min}}$ and $H(r) = H_0(r)-\mu\ell/r$ to leading order in the backreaction. Here $r_{\text{min}}^0$ is the minimum section of $\sigma_t$ in the BTZ geometry. Note that these variations will induce a modification of the momentum $P_t = P_t^0 + \delta P_t$ in \eqref{momentummin}, with $P_t^0 = r_{\text{min}}^0\sqrt{-H_0(r_{\text{min}}^0)}$, namely
	\be\label{variationmomentum}
	\delta P_t^2\, =\, -2r_{\text{min}}^0\,H_0(\sqrt{2}\,r_{\text{min}}^0)\delta r_{\text{min}}\,+\,\mu\ell r_{\text{min}}^0\,.
	\ee
	
	Now that we have all the ingredients, we can expand \eqref{onshellv2} to leading order in the backreaction. Subtracting the zeroth order term to obtain $\delta \text{Vol}(\sigma_t)$ yields
	\be\label{backreactedvol}
    \delta \text{Vol}({\sigma}_t) \, =\, 2\pi\Delta\int_{r^0_{\text{min}}}^{r_\infty}dr\, \dfrac{r^2\left(\mu \ell r\,-\, \delta P^2_t\right)}{(r^2H_0(r)  + (P_t^0)^2)^{3/2}}\;-\,\left.\dfrac{4\pi\Delta r^2}{\sqrt{r^2H(r)+(P^0_t)^2}}\right|_{r_{\text{min}}^0}\delta r_{\text{min}}\,,
	\ee
	where the first term comes from the expansion of the integrand, while the second term arises from the expansion of $r_{\text{min}}$ in the lower limit of the integral. Using \eqref{variationmomentum} and performing the integrals in terms of hypergeometric functions, we see that the divergences of both terms at $r_{\text{min}}^0$ cancel and \eqref{backreactedvol} is finite. We shall write it as the difference between two finite and positive volumes, 
	\begin{equation}
	    \delta \text{Vol}({\sigma}_t) = V_1 -V_2\,, 
	\end{equation}
	with
	\begin{gather}
	V_1 \, =\, 2\pi\Delta\mu\ell\,\int_{r^0_{\text{min}}}^{r_\infty}dr\, \dfrac{r^2(r\,-\,r^0_{\text{min}})}{(r^2H_0(r)  + (P_t^0)^2)^{3/2}}\,,\label{v1}\\
	V_2\, = \,- 4\pi\Delta \,\delta r_{\text{min}}\,\left(\int_{r^0_{\text{min}}}^{r_\infty}dr\, \dfrac{r^2\,r_{\text{min}}^0H_0(\sqrt{2}r_{\text{min}}^0)}{(r^2H_0(r)  + (P_t^0)^2)^{3/2}}\,-\,\left.\dfrac{r^2}{\sqrt{r^2H(r)+(P^0_t)^2}}\right|_{r_{\text{min}}^0}\right)\,\label{v2}.
	\end{gather}
	The integrals can be expressed in terms of hypergeometric and elliptic functions but they are not illuminating. Nevertheless, we can readily understand the main features of this result.
	The positive contribution $V_1$ to \eqref{backreactedvol} comes from the negative change in the redshift factor $H(r)$, which intuitively makes spatial volumes larger. The negative contribution $-V_2$ comes from the fact that the horizon is larger, $r_+ \approx \ell_3+\mu\ell/2$, so there is less space initially. The intuition is that $V_2$ dominates at early times, making the overall backreaction correction negative, but this is reversed at late times when $V_1$ dominates and makes the leading order backreaction positive.

	In order to verify that this interpretation is correct, we will analyze the leading order backreaction for the initial slice $t=0$ explicitly. 
	At $t=0$, the slice $\overline{\sigma}_0$ reaches its minimum section at the horizon, $r_{\text{min}} = r_+$. Also, the slice $\sigma_0$ in the BTZ geometry has $r^0_{\text{min}} = \ell_3$. Therefore, we have $P_t = P_t^0 = \delta P_t =0$ and $\delta r_{\text{min}} = r_+-\ell_3 \approx \mu\ell/2$, so we can integrate \eqref{v1} and \eqref{v2} explicitly. We have
	\begin{gather}
	V_1 \, =\, 2\pi\Delta\mu\ell\,\int_{\ell_3}^{\infty}dr\, \dfrac{(1\,-\,\ell_3/r)}{(H_0(r))^{3/2}}\, = \,\pi\Delta\mu\ell\ell_3(\pi-2)\,,\\
	V_2 \, = \,-2\pi\Delta\mu\ell\ell_3\left(\int_{\ell_3}^\infty \dfrac{dr}{r (H_0(r))^{3/2}}\,-\,\dfrac{1}{\sqrt{H_0(\ell_3)}}\right)\,= \,\pi^2\Delta \mu\ell\ell_3\,.
	\end{gather}
	Then, as we anticipated, the semiclassical backreaction correction ot the volume is negative \footnote{The negativity of $\delta \text{Vol}({\sigma}_0)$ is not incompatible with the results in \cite{Engelhardt:2021mju,Engelhardt:2021kyp} for wormhole states, since the stress-tensor of the CFT$_3$ \eqref{renst} does not satisfy the weak energy condition.}
	\be\label{leadingbrto}
	\delta \text{Vol}({\sigma}_0) \,=\, V_1 -V_2\, =\,-2\pi\mu\ell\ell_3\Delta\,.
	\ee
	
	We might want to view the difference between the volume complexity of the quBTZ and classical BTZ states as a relative complexity between the two states -- perhaps the complexity of formation of quantum backreaction effects. However, this interpretation is not straightforward, since the two states actually belong to different systems, namely a dCFT for quBTZ, and a defect-less CFT for BTZ.
	This means, in particular, that the VC of the quBTZ state \eqref{qc} always includes the contribution $\mathcal{C}_{\text{UV}}$ of the rigid bath, which is absent in the BTZ state.

	Nevertheless, it should be possible to consider states of the dCFT system which contain a BTZ black hole. These will be product states $\ket{\text{BTZ}}\otimes \ket{\Phi}$ between the defect CFT$_2$ and the CFT$_3$ subsystems. We may fine-tune $\ket{\Phi}$ to reproduce the complexity $\mathcal{C}_{\text{UV}}$ of the bath, and then the volume correction in \eqref{leadingbrto} will give the relative complexity between these states.
	
	Relative to the ground state of the dCFT, the quBTZ state is simpler because the degrees of freedom of the defect CFT$_2$ are entangled with the CFT$_3$. On the other hand, to obtain the product state of a BTZ black hole and a the state of the CFT$_3$ bath, it is necessary to break all the entanglement between both subsystems. We may expect that the number of additional simple gates that this requires is $\sim c_3\sim \ell/G_3$ simple gates. The complexity from \eqref{leadingbrto}
	\begin{equation}
	    \frac{\delta \text{Vol}({\sigma}_0)}{8\pi G_3\ell_3}\propto -\frac{\ell}{G_3}\sim -  c_3
	\end{equation}
	indeed has this behavior, and therefore we regard it as measuring (minus) the complexity in the entanglement structure of the quBTZ state. Of course, this identification is highly dependent on the choice of a particular reference state of the dCFT.

	\subsection{Late-time regime}
	
	Let us now analyze the effect of the leading quantum corrections of the CFT$_3$ \eqref{qc} in the late-time regime of the VC of the system. As we have seen, since the rigid bath is static its complexity remains constant, so the late time effects will be controlled by the semiclassical backreaction on the volume \eqref{backreactedvol}.
	
	We shall again consider the extremal slice $\overline{\sigma}_t$ in the quBTZ metric \eqref{qubtz} to all orders in the backreaction of the CFT$_3$ on the brane. The derivative of \eqref{onshellv} with respect to the anchoring time $t$ is given by the Hamilton-Jacobi equation for $\mathcal{L}_V$, which more precisely reads
	\be
	\,\dfrac{d}{dt}\,\text{Vol}(\overline{\sigma}_t) \, =\,2\pi\Delta \,\left(\left.\dfrac{\partial \mathcal{L}_V}{\partial{\dot{t}}}\right|_{r=r_\infty^{L}}\,+\,\left.\dfrac{\partial \mathcal{L}_V}{\partial{\dot{t}}}\right|_{r=r_\infty^{R}}\,\right)\, = \, 4\pi\Delta \,P_t\,,
	\ee
	where $r_\infty^{L,R}$ are the regularized anchoring surfaces for $\overline{\sigma}_t$ on each of the asymptotic boundaries, and the factor of $2$ on the last expression arises from reflection-symmetry of $\overline{\sigma}_t$. Using \eqref{momentum}, we can express the rate in terms of the canonically normalized time-variable, $\bar{t} = t/\Delta$, as
	\be\label{growth}
	\dfrac{d}{d\bar{t}}\,\text{Vol}(\overline{\sigma}_t) \, = \,4\pi\Delta^2 \,r_{\text{min}}\,\sqrt{-H(r_{\text{min}})}\,,
	\ee
	where $r_{\text{min}}$ is the radius of the minimal section of $\overline{\sigma}_t$ in the black hole interior.
	
	At late times, $t\gg \beta$, the extremal slices $\overline{\sigma}_t$ will accumulate over some radial slice $r_0$ in the black hole interior exponentially fast as measured by the asymptotic time variable $t$ (see \cite{Engelhardt:2013tra,Stanford:2014jda,Barbon:2019tuq,Barbon:2020olv,Barbon:2020uux}). This radial accumulation slice is itself extremal, that is, it satisfies
	\be
	\dfrac{d}{dr}\,\left(r\,\sqrt{-H(r)}\right)\Big|_{r_0}  = \, 0\,.
	\ee
	
	In terms of $r_0$, this gives the cubic equation $4(r_0/\ell_3)^3-2(r_0/\ell_3)-\mu\ell/\ell_3 = 0$, which we can solve for the positive root explicitly
	\be\label{r0}
	\dfrac{r_0}{\ell_3} = \dfrac{1}{\sqrt{6}}\left(\gamma \,+\,\dfrac{1}{\gamma}\right)\,,
	\ee
	where 
	\be
	\gamma = \left(\dfrac{3\sqrt{3}}{2\sqrt{2}}\,\dfrac{\mu\ell}{\ell_3}\,+\,\sqrt{-1+\frac{27}{8}\left(\frac{\mu \ell}{\ell_3}\right)^2}\right)^{1/3}\,,
	\ee
	with $\text{Re}(\gamma) > 0$. Note that for $\ell/\ell_3 \rightarrow 0$, we recover $r_0 = \ell_3/\sqrt{2}$, which is precisely the radial position of the accumulation slice for the BTZ black hole. 
	
	The asymptotic late-time linear growth for the extremal volume of $\overline{\sigma}_t$ is then obtained by substituting $r_{\text{min}} \sim r_0$ in \eqref{growth}. Expanding the result to linear order in $\ell/\ell_3$ we get\footnote{Note that the time-derivative of the backreaction is positive, which means that $V_1$ in \eqref{v1} will dominate over $V_2$ in \eqref{v2} for late times $t \gsim \ell_3$, making $\delta \text{Vol}(\sigma_t)$ positive.}
	\be\label{linearvnew}
	\left.\dfrac{d}{d\bar{t}}\,\text{Vol}(\overline{\sigma}_t) \right|_{t\gg \beta}\, \approx \,4\pi\Delta^2 \,r_{0}\,\sqrt{-H(r_{0})}\, = \,2\pi \Delta^2\,\ell_{3}\,\left(1\,+\,\sqrt{2}\,\dfrac{\mu\ell}{\ell_3}\,+\,\dots\right)\,.
	\ee
	
	From the quantum-corrected VC formula \eqref{qc}, this rate will control the late-time growth of the VC of the quBTZ state. With the normalization in \eqref{qc} and using that $\Delta^2 = 8G_3M$ to this order, we arrive to the asymptotic rate of VC
	\be\label{rate}
	\left.\dfrac{d\mathcal{C}_V}{d\bar{t}}\right|_{t\gg \beta}\, = \,2M\,\left(1\,+\,\sqrt{2}\,\dfrac{\mu\ell}{\ell_3}\,+\,\dots\right)\,,
	\ee
	where the dots represent subleading contributions in $\ell/\ell_{3}$. The first term is the late-time VC rate of a BTZ black hole of mass $M$, and the second term is the leading quantum correction to the asymptotic rate. 
	
	To compare this result with the expected slope from the point of view of computational complexity, we shall moreover expand the expected rate $TS_{\text{gen}}$ to linear order 
	\be\label{tsgenqc}
	TS_{\text{gen}}\, = \,2M\,\left(1\,+\,\dfrac{3+x_1^3}{1+x_1^3}\dfrac{\mu\ell}{2\ell_3}\,+\,\dots\right)\,.	
	\ee
	
\begin{figure}[h]
		\centering
		\includegraphics[width = .8\textwidth]{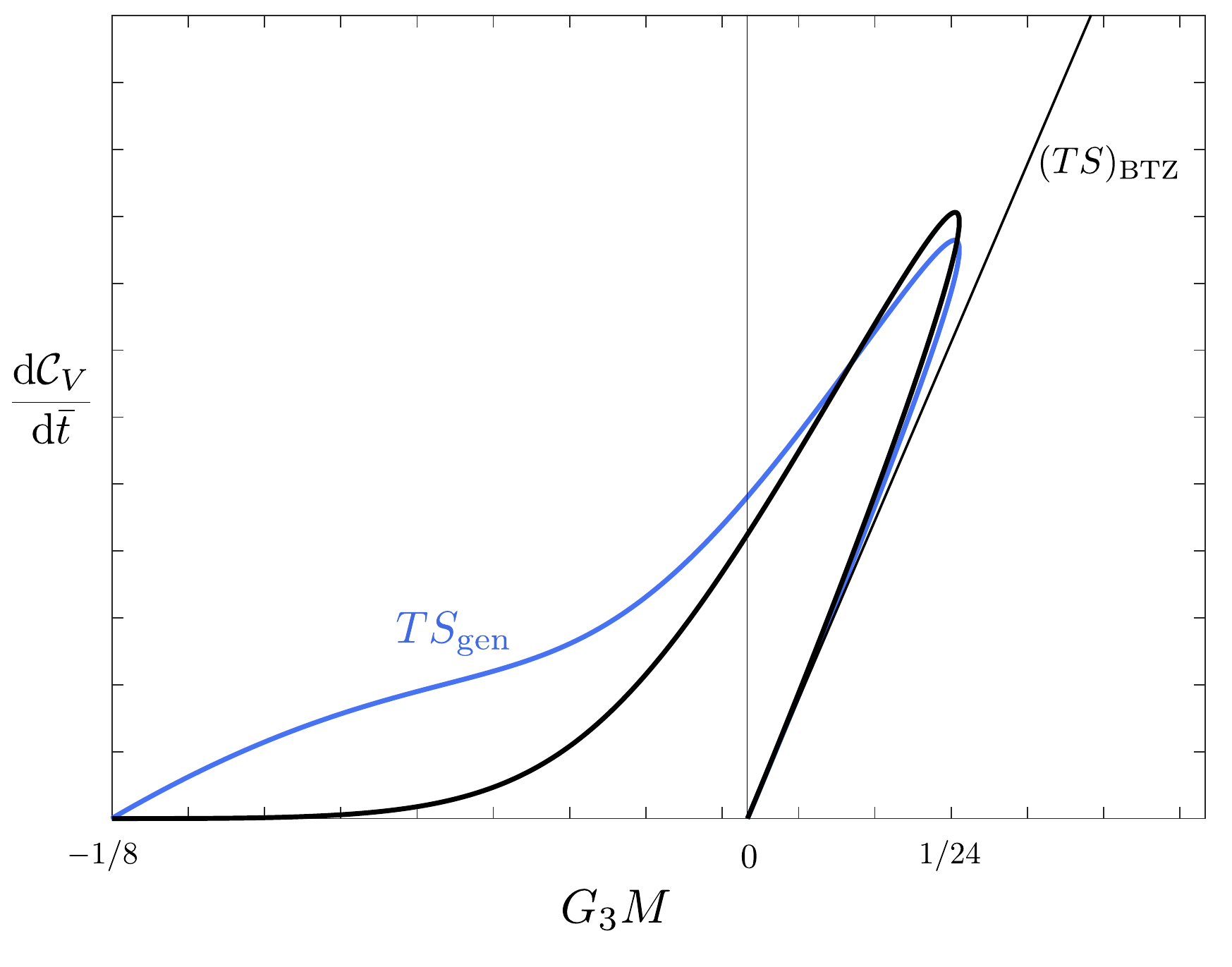}
		\caption{\small The late-time rate of VC for $g_{\text{eff}} = 1/4$. The quBTZ black hole ($M>0$) computes at a faster rate than the BTZ, and it reproduces the behavior of $TS_{\text{gen}}$ up to $\mathcal{O}(1)$ coefficients in the leading quantum-correction. For the branch of quantum-dressed conical defects ($M<0$), the rate of computation is a quantum effect suppressed by $g_{\text{eff}}$ with respect to $TS_{\text{gen}}$.}
		\label{fig::cv}
	\end{figure}

	We plot the comparison between quantum-corrected slope \eqref{rate} to the expected value \eqref{tsgenqc} in Fig.~\ref{fig::cv}. The quantum-correction to \eqref{rate} is positive, which indicates that this system computes faster than the respective BTZ black hole of the same mass $M$. Similarly, the quantum-correction to the expected computation rate $TS_{\text{gen}}$ is also positive. The ratio between the quantum corrections in $d\mathcal{C}_V/d\bar{t}$ and in $TS_{\text{gen}}$ is an $\mathcal{O}(1)$ number (which ranges from $2\sqrt{2}/3$ to $2\sqrt{2}$) that depends on the mass and thermodynamic branch of the black hole. 
	
	\subsubsection*{Quantum-dressed conical defects in AdS$_3$}
	
	We can readily repeat the same analysis for the small AdS$_3$ black holes that appear as quantum-dressed conical defects, which simply require setting $\kappa = +1$ in the AdS C-metric \eqref{cmetric}. Although formally straightforward, an important difference is that in this case the classical solutions do not have horizons and therefore their complexity remains constant. The classical limit $\ell\to 0$ of the quantum-dressed configurations is discontinuous, and the topology of the extremal slice $\Sigma_t$ changes, as it no longer connects two sides. 
	
	The appropriate analogue of \eqref{qc} is then
	\be\label{qcdef}
	\mathcal{C}_V(t)\, =\,\mathcal{C}^{\text{AdS}_3}_V\,+ \,\mathcal{C}_{\text{UV}}\,+\,\dfrac{\delta \text{Vol}(\sigma_t)}{8\pi G_3 \ell_3 }\,+\,\dots\,,
	\ee
	where now the classical part is the (constant in time) VC of two angular defects in AdS$_3$. The growth rate of the extremal volume is again computed by \eqref{growth} with $\kappa = +1$. We find that
	\be\label{ratedef}
	\left.\dfrac{d\mathcal{C}_V}{d\bar{t}}\right|_{t\gg \beta}\, = \,2|M|\,\dfrac{\mu\ell}{\ell_3}\,+\dots\,.
	\ee
	This is not a classical growth rate of VC, but it was expected since these black holes are quantum in origin, with a horizon radius that is $r_+ \propto \ell$. Curiously, since their temperature is $\sim 1/r_+$ (they are small black holes) and their entropy is $\sim r_+$ (they are three-dimensional), in the classical limit, even if the limiting solution is a horizonless conical defect, the product $TS_{\text{gen}}$ takes a non-zero value
	\be\label{tsgendefqc}
	TS_{\text{gen}}\sim 4|M|(M-M_{\text{min}})\,+\,\dots\,,
	\ee
	where $M_{\text{min}} = -(8G_3)^{-1}$ is the mass of empty AdS$_3$. The suppression of \eqref{ratedef} with respect to \eqref{tsgendefqc} is illustrated in Fig.~\ref{fig::cv}.

	\section{Action Complexity}
	\label{sec::AC}
	
	We now turn to examining how the other major proposal for holographic complexity, namely the AC, fares when applied to our quantum-corrected black hole.
	
	In this case, the AC of the time-evolved thermofield double state $\ket{\text{TFD}_\beta(t)}$ of the dCFT system is determined by the on-shell action of the WdW patch in the four-dimensional bulk
	\be\label{ac}
	\mathcal{C}_A(t)\, = \,\dfrac{I(\mathcal{W}_t)}{\pi}\,.
	\ee
	We will adopt the definition of choosing an affine parametrization of the null boundaries of $\mathcal{W}_t$, as well as considering the relevant joint terms at the intersection of the smooth boundaries that render the full action additive (cf. \cite{Hayward:1993my,Lehner:2016vdi}). The action will therefore consist of the Einstein-Hilbert (EH) term with a negative cosmological constant, coupled to a purely tensional brane, with the corresponding boundary terms 
	\begin{gather}
	I(\mathcal{W}_t)\, = \, \dfrac{1}{16\pi G_4}\,\int_{\mathcal{W}_t} \, d^{4}x\,\sqrt{-g}\left(R \,-\,2\Lambda_4\right)\;-\;\dfrac{1}{2\pi G_4\ell} \,\int_{w_t}\,d^{3}y\,\sqrt{-h}  \,\nonumber\\[.4cm] \,\pm\;\dfrac{1}{8\pi G_4}\,\int_{\mathcal{T}, \mathcal{S}}\,d^3y\,\sqrt{|h|}\,K\;+\;\dfrac{1}{8\pi G_4}\,\sum_{i}\,\int_{\mathcal{B}_i}\,d^2\,y \,\sqrt{\eta_i}\,a_i\,,\label{action}
	\end{gather}
	where $w_t = \mathcal{W}_t \cap \textbf{brane}$, $h_{ab}$ is the induced metric on the brane, $\mathcal{T}, \mathcal{S}\subset \partial \mathcal{W}_t$ refer to the timelike/spacelike hypersurfaces of the boundary, $\mathcal{B}_i$ is the intersection between two smooth hypersurfaces, $\eta_i$ is the induced metric on $\mathcal{B}_i$ and $a_i$ is some function that depends on the `boost angle' between the normals to the hypersurfaces (cf. \cite{Hayward:1993my,Lehner:2016vdi,Carmi:2016wjl}).
	
	The WdW patch $\mathcal{W}$ is by definition the bulk causal domain of dependence $D(\Sigma)$ of any spacelike hypersurface $\Sigma$ anchored to the corresponding boundary slice. For spherically symmetric black holes in AdS, $\mathcal{W}_t$ is a `causal diamond' possibly cut by the future/past singularities. In more general situations with less symmetry, such as the AdS C-metric, the shape of $\mathcal{W}_t$ will be generically more complicated due to the formation of caustics.
	
	To simplify the discussion, we will consider a regularized version of the WdW patch, $\widetilde{\mathcal{W}}_t$, defined as $D(\widetilde{\Sigma})$ of any spacelike hypersurface $\widetilde{\Sigma}$ anchored to $r=\infty$ and constant $t$. This new anchoring boundary lies at finite bulk proper distance from the brane, since, as we saw in Sec.~\ref{sec::quBTZ}, the bulk spacetime extends beyond $r=\infty$ into negative values of $r$. It trivially follows that $\widetilde{\mathcal{W}}_t \subset \mathcal{W}_t$. We will regard $\widetilde{\mathcal{W}}_t$ as the relevant object that describes the holographic complexity of the degrees of freedom of the interior of the quBTZ black hole, and in particular we will determine 
	\be\label{irac}
	\widetilde{\mathcal{C}}_A(t)\, = \,\dfrac{I(\widetilde{\mathcal{W}}_t)}{\pi}\,.
	\ee
	From the additivity of the action $\mathcal{C}_A\, = \,\widetilde{\mathcal{C}}_A\,+\,\mathcal{C}_{\text{UV}}$, where the UV part is the action evaluated on $\mathcal{W}_t \setminus \widetilde{\mathcal{W}}_t$. The UV complexity $\mathcal{C}_{\text{UV}}$ is associated to the short-range correlations in the state of the dCFT system. As in the case of VC, from staticity of the spacetime outside the black hole, the UV part of AC gives a constant at late times, and the late-time growth of the AC is controlled by the regularized action \eqref{irac} (see Appendix~\ref{appendix::B} for a more detailed argument).

	\subsection{Regularized WdW patch}
	\label{sec::AC1}
	
	From the properties of the AdS C-metric \eqref{cmetric}, it is possible to see that $\widetilde{\mathcal{W}}_t$ is in fact a `causal diamond'. This is a consequence of the fact that the AdS C-metric is conformal to a (warped) product metric on $\mathcal{M}_2 \times \mathbf{D} $ where $\mathcal{M}_2$ is the $(t,r)$ part of spacetime and $\mathbf{D}$ is the $(x,\phi)$ disk\footnote{With a double-sided brane, two disks $\mathbf{D}$ are glued along the brane giving a topological 2-sphere.}. The conformal factor does not alter the causal structure, and in particular null curves will locally satisfy 
	\be
	H(r)\,\dot{t}^2\, = \, \dfrac{\dot{r}^2}{H(r)} \,+\,r^2\left(\dfrac{\dot{x}^2}{G(x)}\,+\,G(x)\,\dot{\phi}^2\right)\,,
	\ee
	where the dot represents the derivative with respect to an arbitrary parameter along the curve. Since the metric on $\mathbf{D}$ is positive-definite, we can write \footnote{We have implicitly assumed that $H(r)>0$. The inequality in the expression \eqref{ineq} would be reversed for $H(r)<0$, which leads to the same conclusions for radial null geodesics in the black hole interior.}
	\be\label{ineq}
	\left(\dfrac{dt}{dr}\right)^2\, \geq  \, \dfrac{1}{H(r)^2} \,,
	\ee
	and the equality is saturated only if $\dot{x} = \dot{\phi} = 0$. These conditions can indeed be satisfied along radial null geodesics. The condition $\ddot{\phi} = 0$ follows from axial symmetry of the AdS C-metric and, less trivially, $\ddot{x} = 0$ can be consistently imposed using the equations of motion.
	
	The inequality \eqref{ineq} translates into the fact that any null curve that moves along the transverse space $\mathbf{D}$ takes longer time to reach a given radial position. If there is a radial null geodesic $\gamma$ connecting some point $p$ on the regularized boundary at $r=\infty$ to some point $q$ in the bulk, then this will mean that $\gamma$ is indeed the only causal path that connects $p$ and $q$, i.e. all other causal paths take longer time. Moreover,  this radial geodesic $\gamma$ will be the only causal path connecting $q$ to any other point of the regularized boundary. All other points of the regularized boundary have different positions on $\mathbf{D}$ and therefore it is not possible for the corresponding paths connecting them to $q$ to saturate \eqref{ineq}. We reach the conclusion that the boundary of $\widetilde{\mathcal{W}}_t$ is determined by following future/past radial null geodesics thrown from each point at $r=\infty$, and hence $\widetilde{\mathcal{W}}_t$ has the geometric structure of a `causal diamond'.
	
\begin{figure}[h]
		\centering
		\includegraphics[width = .5\textwidth]{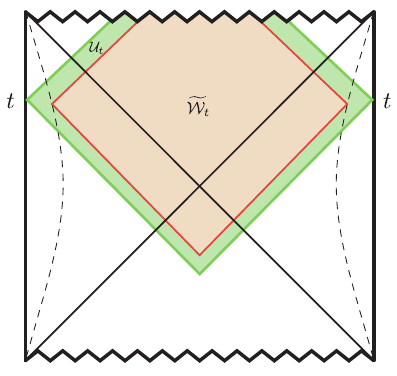}
		\caption{\small The WdW patch $\mathcal{W}_t$ lies in between the red causal diamond $\widetilde{\mathcal{W}}_t$, and the green causal diamond $\mathcal{U}_t$. The figure illustrates a constant $(x,\phi)$ section of the geometry. The dashed lines at finite distance from the horizon correspond to $r=\infty$ in our coordinates.}
		\label{wtildewu}
	\end{figure}

	For later purposes (see Appendix~\ref{appendix::B}), we will also define a new `causal diamond' $\mathcal{U}_t$ by throwing future/past radial null geodesics from the real asymptotic boundary. For any point $q  \notin \mathcal{U}_t$ there is a timelike radial curve connecting $q$ to the boundary slice, which means that $q  \notin {\mathcal{W}}_t$. Thence we have found two diamonds that are tractable to work with and which satisfy $\widetilde{\mathcal{W}}_t\subset \mathcal{W}_t \subset \mathcal{U}_t$ (see Fig.~\ref{wtildewu}). Note that radial geodesics emanating from the asymptotic boundary will not determine the boundary of $\mathcal{W}_t $ because different points of the asymptotic boundary will have different radial positions. The precise characterization of $ {\mathcal{W}}_t$ will involve shortest-time null geodesics connecting bulk points to the asymptotic boundary, and these will in general move along the transverse $\mathbf{D}$.

	\subsubsection*{Quantum onset}
	
	We have seen that the null boundaries of $\widetilde{\mathcal{W}}_t$ are generated by radial null geodesics that saturate the inequality \eqref{ineq}. Therefore, their trajectories will have constant $t \,\pm \, r_*$, where the tortoise coordinate $r_*$ is defined with respect to $r=\infty$ by
	\be\label{tortoise}
	r_*(r)\, = \, \int_{r}^{\infty}\,\dfrac{dr'}{H(r')}\,.
	\ee
	
\begin{figure}[h]
		\centering
		\includegraphics[width = .6\textwidth]{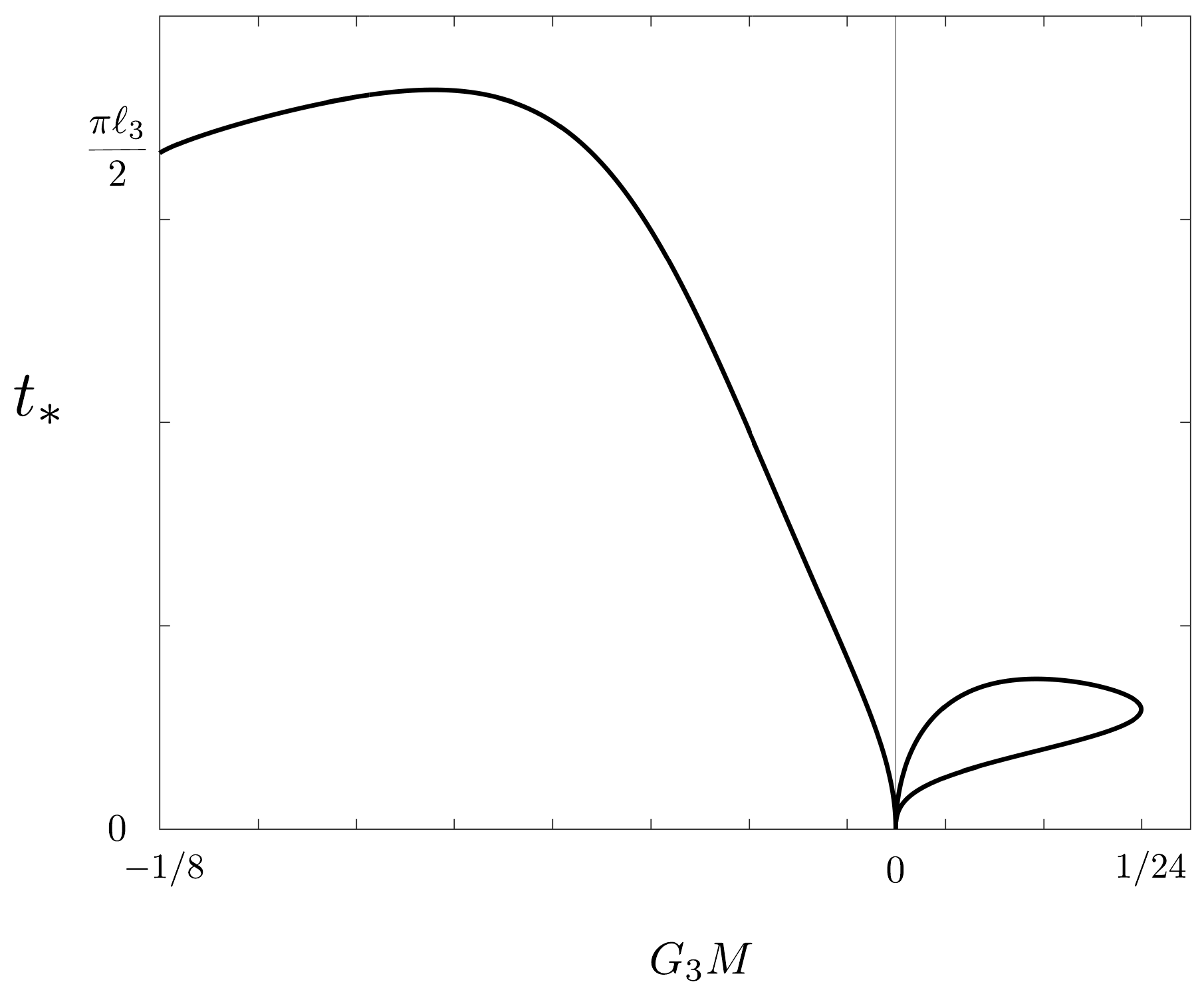}
		\caption{\small Onset of computation $t_*$ for $\ell/\ell_3= 1/3$. For the quBTZ black hole ($M>0$), the delay is a quantum effect suppressed by $\ell/\ell_3$. The delay for small quantum-dressed conical defects ($M<0$) coincides with the onset for small Schwarzschild-AdS black holes. The delay time $t_*$ in this regime becomes a property of the AdS$_3$ `box' (see \protect\cite{Barbon:2017tja}).}
		\label{tstar}
	\end{figure}

	Without loss of generality, we will restrict to the case of forward time-evolution $t \geq 0$ of both asymptotic boundaries. Initially, there will be a period $t \leq t_* = r_*(0)$ for which $\widetilde{\mathcal{C}}_A(t)$ remains constant. The geometric reason behind this period of non-computation is that $\widetilde{\mathcal{W}}_t$ intersects both the past and future singularities, and the effects on the black hole region get compensated by the effects on the white hole region (see \cite{Brown:2015bva,Carmi:2017jqz,Barbon:2017tja}). This is strictly never true for the BTZ black hole, for which $t_*=0$. The onset of computation $t_*$ for the quBTZ is then a purely quantum effect of the CFT$_3$ (see Fig.~\ref{tstar}). We shall assume that the computation has started $t>t_*$ for the rest of the section.

	\subsection{Regularized action}
	\label{sec::ac2}
	
	Since $\widetilde{\mathcal{W}}_t$ has the geometric structure a `causal diamond', computing the terms of the on-shell action \eqref{ac} becomes straightforward. The regularized patch $\widetilde{\mathcal{W}}_t$ does in fact coincide with the WdW patch ${\mathcal{W}}_t$ along the brane, i.e. $\widetilde{w}_t =w_t$. 
	
	\subsubsection*{Bulk contribution}
	
	The purely bulk contribution $I_{\text{bulk}}(\widetilde{\mathcal{W}}_t)$ to the action \eqref{irac} will consist of the EH action $I_{\text{EH}}(\widetilde{\mathcal{W}}_t)$ plus the bare brane action $I_{\text{brane}}(\widetilde{w}_t)$. Since \eqref{cmetric} is locally a vacuum solution away from the brane
	\be
	R - 2 \Lambda_4 = -6/\ell_4^2\;\hspace{.7cm}\text{for}\hspace{.3cm} x\neq 0\,,
	\ee
	and the EH action then gives a contribution which is proportional to the spacetime volume of $\widetilde{\mathcal{W}}_t$. Additionally, the brane will contribute to this term since it affects the curvature via Israel's junction conditions
	\be
	R_{\mu\nu}\,-\,\dfrac{1}{2}\,R\,g_{\mu\nu}+\,\Lambda_4\,g_{\mu\nu} = -\dfrac{4}{\ell}\, h_{\mu\nu}\, \delta(X)\,,
	\ee
	where $X$ is a normal parameter to the brane that grows towards $x>0$, and $h_{\mu\nu}$ is the induced metric on the brane. The normal parameter $X$ satisfies ${dX}/{dx} = 1/r$ at $x = 0$, and the integrand then becomes $R - 2 \Lambda_4 = -6/\ell_4^2 \,+\, {12\,\delta(x)}/{r\ell}$. The delta contribution then collapses the EH action into an volume integral over $w_t$. The full EH action is then
	\be\label{eh}
	I_{\text{EH}}(\widetilde{\mathcal{W}}_t)\;= \; -\dfrac{3\,\text{Vol}(\widetilde{\mathcal{W}}_t)}{8\pi G_4 \ell_4^2}\,+\,\dfrac{3\,\text{Vol}(w_t)}{4\pi G_4 \ell}\,.
	\ee
	Incorporating the bare brane contribution results in the total bulk contribution
	\be\label{ibulk}
	I_{\text{bulk}}(\mathcal{W}_t)\, = \,-\dfrac{3\,\text{Vol}(\widetilde{\mathcal{W}}_t)}{8\pi G_4 \ell_4^2}\,+\,\dfrac{\text{Vol}(w_t)}{4\pi G_4\ell}\,.
	\ee
	
	\subsubsection*{Spacetime volume}
	
\begin{figure}[h]
		\centering
		\includegraphics[width = .5\textwidth]{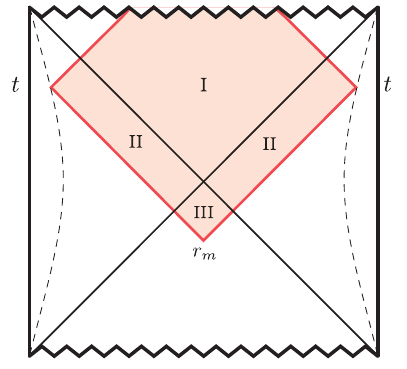}
		\caption{\small Regularized WdW patch $\widetilde{\mathcal{W}}_t$ for $t>t_*$. The radius of the lower tip $r_m(t)$ is implicitly determined by the equation $t = r_*(r_m)$.}
		\label{wdw2div}
	\end{figure}

	We will now compute $\text{Vol}(\widetilde{\mathcal{W}}_t)$ and $\text{Vol}(w_t)$ entering in \eqref{ibulk}. The regularized WdW patch will intersect the future singularity and the lower tip of $\widetilde{\mathcal{W}}_t$ will be localized at $r_m(t)>0$ in the white hole region. A systematic way to compute the volume of $\widetilde{\mathcal{W}}_t$  is to first divide it into the different regions shown in Fig.~\ref{wdw2div}. Performing the corresponding integration over the $t$-coordinate for each of the regions yields
	\begin{gather}
	\text{Vol}(\text{I})\,= \,\int_{0}^{2\pi\Delta}d\phi\,\int_{0}^{x_1}dx\,\int_{0}^{r_+}\,dr\,\dfrac{4\ell^4r^2}{(\ell + xr)^4}\,\left(t+r_*(r)\right)\,,\\
	\text{Vol}(\text{II})\,= \,\int_{0}^{2\pi\Delta}d\phi\,\int_{0}^{x_1}dx\,\int_{r_+}^{\infty}\,dr\,\dfrac{8\ell^4r^2}{(\ell + xr)^4}\,r_*(r)\,,\\
	\text{Vol}(\text{III})\,= \,\int_{0}^{2\pi\Delta}d\phi\,\int_{0}^{x_1}dx\,\int_{r_m}^{r_+}\,dr\,\dfrac{4\ell^4r^2}{(\ell + xr)^4}\,\left(-t+r_*(r)\right)\,,
	\end{gather}
	where we have added a factor of $2$ to account for the two-sided spacetime $x<0$.
	
	Adding the three contributions to the volume of $\mathcal{W}_t$, we obtain
	\be\label{svc}
	\text{Vol}(\widetilde{\mathcal{W}}_t) = \, \text{V}_0\,+\,8\pi\Delta \ell^4\,\int_{0}^{x_1}dx\,\int_{0}^{r_m}\,dr\,\dfrac{r^2}{(\ell + xr)^4}\,\left(t-r_*(r)\right)\,,
	\ee
	where $\text{V}_0$ is the constant volume
	\be
	\text{V}_0\,= \,{16\pi\Delta \ell^4}\,\int_{0}^{x_1}dx\,\int_{0}^{\infty}\,dr\,\dfrac{r^2}{(\ell + xr)^4}\,r_*(r)\,.
	\ee
	
	The contribution from the bulk ($x\neq 0$) to all of the quantities defined so far for $\widetilde{\mathcal{W}}_t$ is finite, namely because the anchoring surface at $r = \infty$ lies at finite proper distance from the horizon of the black hole. For the case of the intersection with the brane, however, an extra radial cutoff is strictly required at $r =r_\infty <\infty$ to contemplate the intersection between the regularized boundary and the asymptotic boundary \footnote{For a careful treatment of the regularization of the WdW patch in a Karch-Randall setup, see \cite{Chapman:2018bqj,Braccia:2019xxi}.}.

	The volume of $w_t$ is given by repeating the previous analysis for the case of the brane. The result is
	\be\label{svcbrane}
	\text{Vol}(w_t) = \, \text{v}_0\,+\,4\pi\Delta\,\int_{0}^{r_m}\,dr\,r\,\left(t-r_*(r)\right)\,,
	\ee
	where $\text{v}_0$ is the constant volume
	\be
	\text{v}_0\,= \,{8\pi\Delta \ell^4}\,\int_{0}^{\infty}\,dr\,r\,r_*(r)\,.
	\ee

	\subsubsection*{GHY contribution}
	
	For $t>t_*$, the regularized WdW patch $\widetilde{\mathcal{W}}_t$ will intersect the future singularity. Therefore, the action \eqref{ac} will include a Gibbons-Hawking-York (GHY) contribution $I_{\text{GHY}}(\widetilde{\mathcal{W}}_t)$ coming from the regularized future singularity at $r=\epsilon$. The future-pointing unit-normal to this surface is $n_\mu \propto - (dr)_\mu$ with a normalization given by
	\be\label{normal}
	n_{\mu}\, = \,- \dfrac{\ell}{\ell + xr}\,\dfrac{(dr)_{\mu}}{\sqrt{-H}}\,.
	\ee
	
	The integrand of the GHY term is the extrinsic curvature $K$ of the future singularity
	\begin{equation}\label{extrcurv}
	K = \dfrac{1}{\sqrt{-g}}\,\partial_\mu \left(\sqrt{-g} \;n^\mu\right)\, = \,\dfrac{\sqrt{-H}(\ell + xr)}{2\,\ell}\,\left(\dfrac{\partial_r \,H}{H}\,-\,\dfrac{6x}{\ell + xr}\,+\,\dfrac{4}{r}\right)\,,
	\end{equation}
	evaluated at $r = \epsilon$.
	
	The intersection between $\widetilde{\mathcal{W}}_t$ and the regularized future singularity extends in the $t$ direction along the interval $[-t_{\text{s}},t_{\text{s}}]$, where $t_{\text{s}} = t+r_*(\epsilon)$. Considering a factor of $2$ to account for the two sides of the system, the GHY contribution is then
	\begin{gather}
	I_{\text{GHY}}(\widetilde{\mathcal{W}}_t)\, = \,\lim\limits_{\epsilon\rightarrow 0} \,\dfrac{2}{8\pi G_4}\,\int_0^{2\pi \Delta}\,d\phi \int_0^{x_1}dx\,\int_{-t_{\text{s}}}^{t_{\text{s}}}\,dt'\,\sqrt{-H}\,\dfrac{\ell^3\,\epsilon^2}{(\ell+x\epsilon)^3}\;K\,\nonumber \\[.4cm]
	= \, \dfrac{3\Delta\,x_1\,\mu\,\ell}{2\, G_4}\,\left(t+r_*(0)\right)\,. \hspace{3.5cm}\label{ighy}
	\end{gather}

	\subsubsection*{Contributions from joints}
	
\begin{figure}[!htbp]
		\begin{center}
			\begin{subfigure}{0.45\linewidth}
				\centering
				\includegraphics[width=1\textwidth]{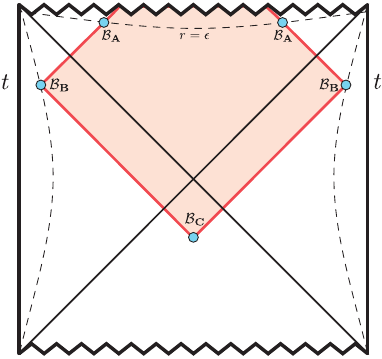}\quad
			\end{subfigure}
			\hfill
			\begin{subfigure}{0.45\linewidth}
				\centering
				\includegraphics[width=1\textwidth]{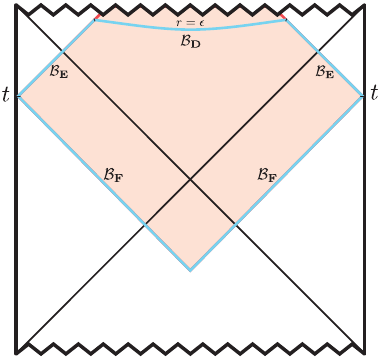}\quad
			\end{subfigure}
		\end{center}
		\caption{\small  \small \textbf{Left}: The `standard joint surfaces' in the bulk for a constant $(x,\phi)$ section of the geometry. All of the blue points represent two-dimensional surfaces that extend in the $x$ and $\phi$ directions. \textbf{Right}: The `new joint surfaces' for a constant $\phi$ section of the brane. The blue lines represent two-dimensional surfaces that extend in the $\phi$ direction along the brane.}
		\label{fig:jointwdw}
	\end{figure}

	For the regularized WdW patch $\widetilde{\mathcal{W}}_t$, there will be two classes of joint surfaces which \textit{a priori} are relevant for the joint contribution $I_{\text{joint}}(\widetilde{\mathcal{W}}_t)$ to the action \eqref{irac} (see Fig.~\ref{fig:jointwdw}). The `standard joint surfaces' lie in the four dimensional bulk, and correspond to:
	\begin{itemize}
		\item[-] {$\mathcal{B}_{\textbf{A}}$}: The intersection of the future null boundary of $\widetilde{\mathcal{W}}_t$ with the future singularity. 
		\item[-] {$\mathcal{B}_{\textbf{B}}$}: The intersection of the future and past null boundaries of $\widetilde{\mathcal{W}}_t$ at $r = \infty$.
		\item[-] {$\mathcal{B}_{\textbf{C}}$}:  Caustics of the past null boundary of $\widetilde{\mathcal{W}}_t$ in the white hole region at $r = r_m(t)$.
	\end{itemize}

	Moreover, the presence of the localized vacuum energy of the brane at $x=0$ makes the spacetime non-smooth across it, and therefore $\widetilde{\mathcal{W}}_t$ will inherit this non-smoothness when it intersects the brane. The `new joint surfaces' will be therefore localized on the brane, and will correspond to:
	\begin{itemize}
		\item[-] {$\mathcal{B}_{\textbf{D}}$}: The intersection between the two sides of the regularized future singularity at $r=\epsilon$.
		\item[-] {$\mathcal{B}_{\textbf{E}}$}: The intersection between the two sides of the future null boundary.
		\item[-] {$\mathcal{B}_{\textbf{F}}$}: The intersection between the two sides of the past null boundary.
	\end{itemize}

	The surface $\mathcal{B}_{\textbf{A}}$ arises as the intersection between a null and a spacelike hypersurface, with outward-directed normals $n^{(1)}_\mu = \alpha_1\left[(dt)_\mu-(dr_*)_\mu\right]$ and $n^{(2)}_\mu$ in \eqref{normal}, respectively. The constant $\alpha_1>0$ specifies the parametrization amongst all possible future-directed affine parametrizations of the null generators. The `additivity rules' of \cite{Hayward:1993my,Lehner:2016vdi} prescribe the integrand to be $a_{\textbf{A}}\, = \, -\log \left|{n^{(1)}\cdot n^{(2)}}\right|$. Taking the limit $\epsilon \rightarrow 0$, the total contribution of this surface vanishes
	\be
	I^{\textbf{A}}_{\text{joint}}(\widetilde{\mathcal{W}}_t)\, = \,\lim\limits_{\epsilon\rightarrow 0}\,\dfrac{2\times 2}{8\pi G_4}\,\int_{0}^{2\pi\Delta}d\phi\,\int_{0}^{x_1}dx\,\dfrac{\ell^2 \epsilon^2}{(\ell + x\epsilon)^2}\;\log \left(\alpha_1\,\dfrac{(\ell + x\epsilon)}{\ell\,\sqrt{-H(\epsilon)}}\,\right)\;= \, 0\,.
	\ee
	
	The next surface on the list, $\mathcal{B}_{\textbf{B}}$, is the result of the intersection between the future and past null boundaries at $r=\infty$. The outward-directed normals are $n^{(1)}_\mu =\alpha_1\left[(dt)_\mu-(dr_*)_\mu\right]$ and $n^{(2)}_\mu = -\alpha_2\left[(dt)_\mu+(dr_*)_\mu\right]$, with $\alpha_1,\alpha_2 >0$. The prescription in this case is to take $a_{\textbf{B}}\, = \, -\log ({n^{(1)}\cdot n^{(2)}/2})$. The total contribution of this surface is then
	\be
	I^{\textbf{B}}_{\text{joint}}(\widetilde{\mathcal{W}}_t)\, = \,\lim\limits_{r\rightarrow \infty}\,\dfrac{2\times 2}{8\pi G_4}\,\int_{0}^{2\pi\Delta}d\phi\,\int_{0}^{x_1}dx\,\dfrac{\ell^2 r^2}{(\ell + xr)^2}\;\log \left(\dfrac{\ell^2 H}{(\ell+xr)^2\alpha_1\alpha_2}\,\right)\,.
	\ee
	This term is finite in the bulk but still has a remaining divergence at $x\rightarrow 0$ from the intersection of $r=\infty$ with the asymptotic boundary at the anchoring points of the brane. This divergences can be dealt with by using standard counterterms on the brane after placing a radial cutoff at finite distance $r_\infty < \infty $.
	
	The last surface amongst the `standard' ones, $\mathcal{B}_{\textbf{C}}$, originates as the intersection of the left and right past null boundaries. The corresponding outward-pointing normals are $n^{(1)}_\mu = \alpha_1\,\left[(dt)_\mu-(dr_*)_\mu\right]$ and $n^{(2)}_\mu = -\alpha_2\,\left[(dt)_\mu+(dr_*)_\mu\right]$, for $\alpha_1,\alpha_2 > 0$. The prescription is to integrate $a_{\textbf{C}}\, = \, \log \left|{n^{(1)}\cdot n^{(2)}/2}\right|$, which yields
	\begin{gather}
	I^{\textbf{C}}_{\text{joint}}(\widetilde{\mathcal{W}}_t)\, = \,-\dfrac{2}{8\pi G_4}\,\int_{0}^{2\pi\Delta}d\phi\,\int_{0}^{x_1}dx\,\dfrac{\ell^2 r_m^2}{(\ell + xr_m)^2}\,\log \left(\dfrac{\ell^2}{(\ell+xr_m)^2}\,\dfrac{|H(r_m)|}{\alpha_1\alpha_2}\right)\,\nonumber\\[.4cm] \;= -\dfrac{\Delta\,\ell\, r_m}{2G_4(\ell+x_1r_m)}\left[r_mx_1\left(2+\log \left(\dfrac{|H(r_m)|}{\alpha_1\alpha_2}\right)\right)-2\ell \log \left(\dfrac{\ell}{\ell+x_1r_m}\right)\right]\,.\label{jointc}
	\end{gather}
	
	As for the `new joint surfaces', it turns out that they will yield no contribution to the joint term in the action. The reason is that for all of the two-sided hypersurfaces that intersect the brane and produce these joint surfaces, their normal vector becomes tangent to the brane on both sides of the brane. From the junction conditions, the tangent properties of the spacetime remain continuous across the brane, and so will these normal vectors. In other words, the hypersurfaces will be at least one-time differentiable across the brane. The joint term for such situations naturally vanishes.
	
	For instance, consider $\mathcal{B}_{\textbf{D}}$, which is the result of the intersection of the $r=\epsilon$ hypersurface with the brane. The normals will be radial $n^\pm_\mu \propto -(dr)_\mu$ where $\pm$ refers to the sign of $x$ on each side of the hypersurface. These normal vectors $n_\pm^\mu$ are orthogonal to $(dx)_\mu$, and hence they are tangent to the brane when evaluated at $x= 0^\pm$. The normalization \eqref{normal} for this tangent vectors is a property of the induced metric on the brane, which is continuous. Therefore, the normal vector itself is continuous $n_+^\mu = n_-^\mu$ at $x=0$. The `boost angle' in the integrand of the corresponding joint term $a_{\textbf{D}}\propto\cosh^{-1} (-n_+\cdot n_-) $ will vanish, giving no contribution to $I_{\text{joint}}(\widetilde{\mathcal{W}}_t)$. The same effect happens for the cases of $\mathcal{B}_{\textbf{E}}$ and $\mathcal{B}_{\textbf{F}}$, where the normals are now $n_\mu \propto H(dt)_\mu\,-\,(dr)_\mu$ and $n_\mu \propto H(dt)_\mu\,+\,(dr)_\mu$, respectively.

	\subsection{Time-dependence}

	The time derivative of \eqref{svc} with respect to the upper limit of the radial integral identically vanishes since $t-r_*(r_m) = 0$. The time-dependence of $\text{Vol}(\widetilde{\mathcal{W}}_t)$ and $\text{Vol}(w_t) $ comes then entirely from the respective integrands. Referring to the properly normalized time-variable on the defect, $\bar{t} =  t/\Delta$, we obtain
	\be \label{svcgrowth}
	\dfrac{d}{d\bar{t}}\,\text{Vol}(\widetilde{\mathcal{W}}_t) \,= \,8\pi\Delta^2 \ell^4\,\int_{0}^{x_1}dx\,\int_{0}^{r_m}dr\,\dfrac{r^2}{(\ell + xr)^4}\,\,= \, \dfrac{4\pi \Delta^2\,r_m^2\,\ell}{3}\left(1-\dfrac{\ell^2}{(\ell+x_1r_m)^2}\right)\,,
	\ee
	and
	\be \label{svcgrowthbrae}
	\dfrac{d}{d\bar{t}}\,\text{Vol}(w_t) \,= \,4\pi\Delta^2 \,\int_{0}^{r_m}dr\,r\,\,= \, 2\pi\Delta^2r_m^2\,.
	\ee

	The time-dependence of the bulk contribution \eqref{ibulk} can be directly obtained from these two expressions
	\be\label{bulktder}
	\dfrac{d}{d\bar{t}}\,I_{\text{bulk}}(\widetilde{\mathcal{W}}_t)\,=  \,-\dfrac{\Delta^2\,r_m^2\,\ell}{2G_4\ell_4^2}\,\left(1\,-\,\dfrac{\ell^2}{(\ell+x_1r_m)^2}\right)\;+\;\dfrac{\Delta^2\,r_m^2}{2G_4\ell}\,.
	\ee

	The GHY term \eqref{ighy} grows linearly in time
	\be\label{ghytder}
	\dfrac{dI_{\text{GHY}}(\widetilde{\mathcal{W}}_t)}{d\bar{t}}\,= \,\dfrac{3\Delta^2\, x_1\,\mu\,\ell }{2G_4}\,\,.
	\ee

	For the joint contributions, the only time-dependent contribution comes from $\mathcal{B}_{\textbf{C}}$ in \eqref{jointc}. The speed at which the lower tip of $\widetilde{\mathcal{W}}_t$ approaches the horizon is $dr_m/dt =  -H(r_m)$, so this terms gives
	\begin{gather}
	\dfrac{d}{d\bar{t}}\,I_{\text{joint}}(\widetilde{\mathcal{W}}_t)\,= \,\dfrac{\Delta^2 \ell x_1r_mH(r_m) }{2G_4 (\ell+x_1r_m) }\,\left[4+2\log \left(\dfrac{H(r_m)}{\alpha_1\alpha_2}\right)+\dfrac{r_m H'(r_m)}{H(r_m)}\,+\,\right.\nonumber\\[.4cm] \left.+\,\dfrac{x_1r_m}{\ell+x_1r_m}\left(2\ell + 2\ell \log \left(\dfrac{\ell}{\ell+x_1r_m}\right)-2x_1r_m-x_1r_m\log\left(\dfrac{H(r_m)}{\alpha_1\alpha_2}\right)\right) \right]\,,\label{jointtder}
	\end{gather}
	which depends on the choice of affine parametrization through the positive constants $\alpha_1$ and $\alpha_2$. 
	
	\subsection{Late-time regime}
	
	The lower tip of $\widetilde{\mathcal{W}}_t$ moves towards the bifurcation surface with a velocity given by $dr_m/dt = -H(r_m)$, or with the implicit trajectory $r_*(r_m) = t$. At late times compared to the inverse temperature of the black hole, $t\gg \beta$, the expansion of the trajectory around $r_m\approx r_+$ shows that it approaches the horizon exponentially fast
	\be\label{latetimerm}
	r_+-r_m(t) \sim \exp\left(-\frac{4\pi t}{\Delta\,\beta}\right)\,,
	\ee
	and thence, for the late-time regime $t\gg \beta$, it is possible to consistently assume that $r_m\sim r_+$.
	
	At late times, the expression for the joint contribution \eqref{jointtder} simplifies drastically
	\be\label{jointasympt}
	\dfrac{d}{d\bar{t}}\,I_{\text{joint}} (\widetilde{\mathcal{W}}_t)\, \approx \,\dfrac{\Delta^2 \ell x_1r_+^2H'(r_+) }{2 G_4 (\ell+x_1r_+) }\, = \, 2\,TS_{\text{gen}}\,.
	\ee
	
\begin{figure}[h]
		\centering
		\includegraphics[width = .8\textwidth]{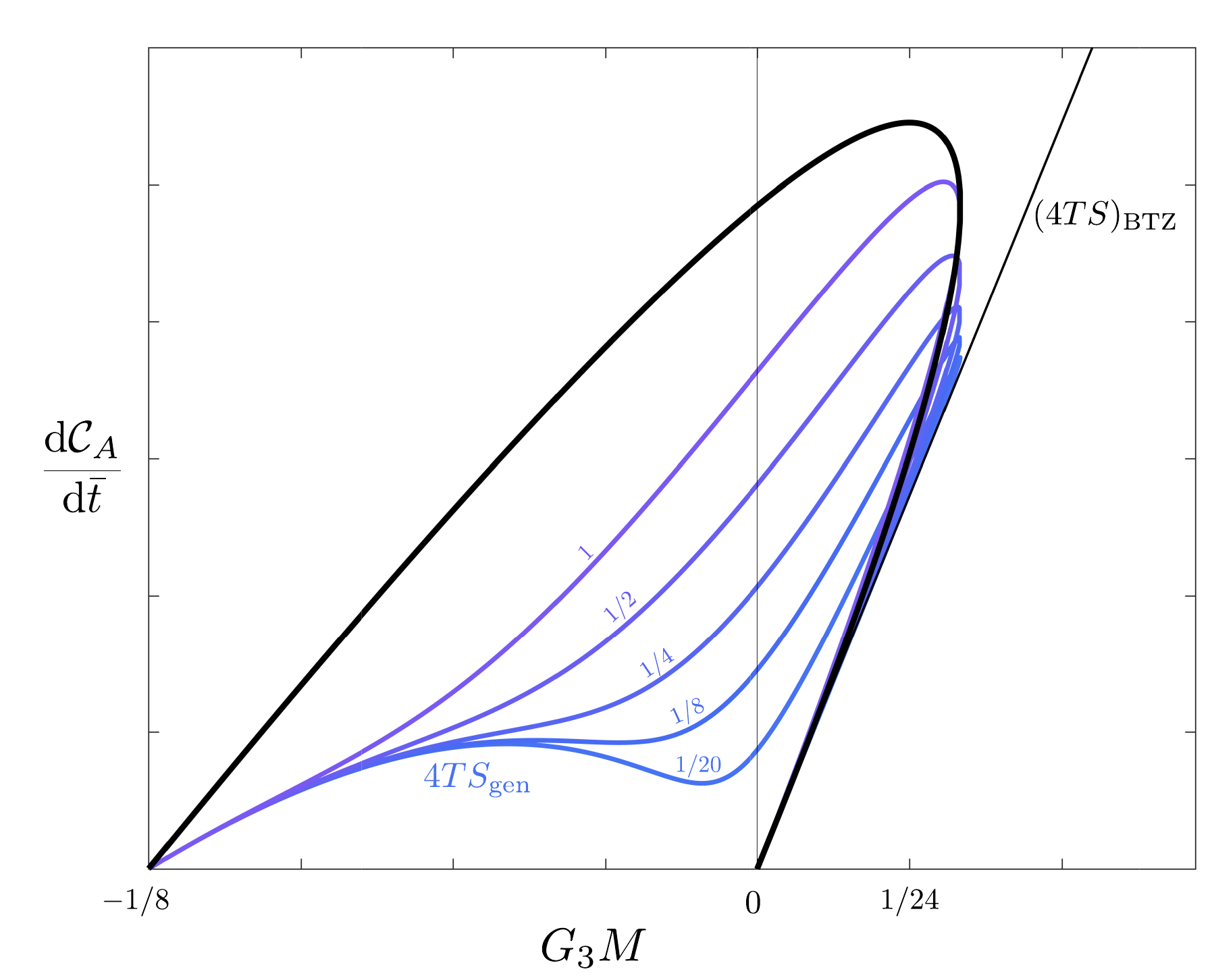}
		\caption{\small The late-time slope of AC, in black, is independent of $g_{\text{eff}}$. This is true both in the quBTZ branch ($M>0$) and in the branch of quantum-dressed conical defects ($M<0$).  In faded blue, the value of $4\,TS_{\text{gen}}$ for the values $g_{\text{eff}} \in \lbrace  1, \frac{1}{2}, \frac{1}{4},\frac{1}{8}, \frac{1}{20} \rbrace $. }
		\label{fig::ca}
	\end{figure}

	Adding the three late time contributions \eqref{bulktder}, \eqref{ghytder} and \eqref{jointasympt}, and using the relations between the parameters \eqref{Deltax1} and \eqref{mux1}, we obtain a remarkably simple expression for the asymptotic rate of AC,
	\be\label{acasympt}
	\left.\dfrac{d{\mathcal{C}_A}}{d\bar{t}}\right|_{t\gg \beta}\, =  \,\dfrac{8M\,x_1\,\mu}{\pi}\,.
	\ee
    Surprisingly, for a given mass $M$ this growth rate is independent of $g_{\text{eff}}$: according to \eqref{MDelta}, \eqref{Deltax1}, and \eqref{mux1}, the functions $x_1(M)$ and $\mu(M)$ are independent of $\ell$, and hence of the position of the brane in the bulk\footnote{For this to be strictly true beyond the linear order in $\ell$, one must keep fixed not $G_3$, but its `renormalized' value $\mathcal{G}_3$ (due to higher-curvature terms) \cite{Emparan:2020znc}. The dependence on the brane position that this may induce in  \eqref{acasympt} is mild and very restricted.}.
	Therefore, the result  \eqref{acasympt} is valid to all orders in $g_{\text{eff}} \sim \ell/\ell_3$ (in the limit $c_3\rightarrow \infty$), and hence, unlike $TS_{\text{gen}}$, the AC does \textit{not} admit a proper semiclassical expansion in powers of $g_{\text{eff}}$ (see Fig.~\ref{fig::ca}).
	
	For the branch of quBTZ black holes with small masses, $G_3M \ll 1$, we can expand $x_1\mu$ in terms of the mass of the black hole as
	\be\label{acdeflim}
	\left.\dfrac{d\mathcal{C}_A}{d\bar{t}}\right|_{t\gg \beta}\;=\,\dfrac{8M}{\pi}\left(1\,+\,2{G}_3 M\,+\,\mathcal{O}({G}_3 M)^2\right)\,.
	\ee
	The AC rate \eqref{acdeflim} does not reduce to the expected complexity growth in the `classical limit' $g_{\text{eff}} \rightarrow 0$. Note that, in this limit, $c_3/c_2 \rightarrow 0$, and the holographic system consists of a pair of two-dimensional conformal defects describing a BTZ black hole. Nevertheless, \eqref{acdeflim} is independent of $g_{\text{eff}}$ and its value does not give the AC rate of a BTZ black hole (see Fig.~\ref{fig::ca}). 
	
	Since \eqref{acasympt} does not depend on the position (or tension) of the brane $\ell/\ell_3$ in the four-dimensional bulk, we can fix the mass $M$ and take the `tensionless limit' $\ell \rightarrow \infty$ to evaluate it---bearing in mind that the effective three-dimensional description is far from being approximately valid. The AdS C-metric \eqref{cmetric} in this limit becomes the metric of a hyperbolic AdS$_4$ black hole, with the brane disappearing and only leaving a mere reflection symmetry condition at the equatorial section. The AC rate  \eqref{acasympt} does match the correct slope for such a four-dimensional black hole
	\be\label{int}
	\left.\dfrac{d\mathcal{C}_A}{d\bar{t}}\right|_{t\gg \beta}\, =\,\dfrac{8Mx_1\mu}{\pi}\, = \, \dfrac{4}{\pi}\,\overline{M}_{\text{ADM}}\,,
	\ee
	where $\overline{M}_{\text{ADM}}$ is the four-dimensional ADM mass (with respect to $\partial/\partial {\bar{t}}$) of the hyperbolic black hole with transverse volume $V_2 = 4\pi \Delta x_1$
	\be\label{admmass}
	\overline{M}_{\text{ADM}} = \dfrac{\Delta V_2\,\mu\ell}{8\pi G_4}\, = \dfrac{\Delta^2 x_1 \mu \ell}{2G_4}\, = \,  \, 2Mx_1\mu \,.
	\ee
	
	\subsubsection*{Quantum-dressed conical defects in AdS$_3$}
	
	For quantum-dressed conical defects, the calculation of AC can be directly extended from the previous results using the $\kappa = +1$ version of the AdS C-metric \eqref{cmetric}. The late-time regime is the same as \eqref{acasympt} with an absolute value on the mass 
	\be\label{acasymptcd}
	\left.\dfrac{d{\mathcal{C}_A}}{d\bar{t}}\right|_{t\gg \beta}\, =  \,\dfrac{8|M|\,x_1\,\mu}{\pi}\,,
	\ee
	which has been incorporated into Fig. \ref{fig::ca} as the $M<0$ branch.
	
	The inconsistency of this late-time rate of AC and a putative quantum-corrected AC formula $\mathcal{C}_A(t) = \mathcal{C}^{\text{AdS}_3}_A\,+\,\delta \mathcal{C}_A(t)$ is perhaps clearer in this case: in the classical limit $\ell/\ell_3 \rightarrow 0$, \eqref{acasymptcd} yields a non-vanishing AC rate for this system, which clashes with the zero computation rate that the static conical defects in AdS$_3$ would have.

	\subsection{Summary}
	
	The most important conclusion of this lengthy analysis of the AC of the thermal state of the dCFT, is that its late-time behavior, computed using the doubly-holographic four-dimensional bulk solution, is independent of the semiclassical expansion parameter $g_{\text{eff}}$ of the effective three-dimensional theory.
	
	This would not be a problem if the value of AC reproduced the classical one for the BTZ black hole with mass $M$---in that case, the result would imply the absence of three-dimensional bulk quantum corrections to AC. But this is not what happens here: \eqref{acasympt} differs parametrically from the late-time slope of AC for BTZ. That is, the AC of this system does not even reproduce the classical term at $\mathcal{O}(g_{\text{eff}}^0)$ for the BTZ black hole. In this regard, AC differs from the behavior of $TS_{\text{gen}}$, which has the expected limit when $g_{\text{eff}}\to 0$.
	
	Instead, what the result \eqref{acasympt} correctly gives is the AC rate \eqref{int} at the opposite limit $g_{\text{eff}}\rightarrow \infty$, where the four-dimensional bulk solution is a hyperbolic AdS$_4$ black hole (or spherical, when $\kappa=+1$). In this limit, the brane becomes tensionless and the defect in the CFT$_3$ disappears. 
	
	
	This discontinuity in the classical limit is puzzling, but its origin can be traced back to the sensitivity of AC to the singularity in the black hole interior; in contrast, VC does not probe the singularity since the extremal spacelike slices always remain away from it. We can regard quantum effects as entering in two ways: first, the backreaction of quantum fields turns the mild singularity at $r=0$ into a strong curvature singularity. Second, quantum fields are strong near this singularity. Their stress tensor behaves as $\sim 1/r^3$, yielding a divergence in the action of the CFT$_3$,\footnote{This can be obtained from the holographic stress tensor as $\delta I_{CFT}=\int T_{\mu\nu}\delta h^{\mu\nu}$.} which cancels (by the equations of motion) against a divergence of the effective three-dimensional gravitational action. The total action is finite, as the result \eqref{acasympt} shows, but there remains a contribution to the AC of the quantum fields that is comparable to the AC of the classical BTZ black hole, and which stays finite as $g_\textrm{eff}\to 0$.
	
	Therefore, it is the sensitivity of AC to the large quantum effects near the black hole singularity that is responsible for the discontinuity that we have found.

	\section{Conclusions and outlook}
	\label{sec:discuss}
	
	We have analyzed the bulk quantum corrections to the VC and AC prescriptions for a semiclassical black hole in AdS$_3$ using holographic conformal fields in the three-dimensional bulk. The AdS$_3$ system we have studied is the quBTZ black hole of \cite{Emparan:2020znc}, a semiclassical black hole solution of a three dimensional gravitational theory with holographic CFT$_3$ as matter, which is coupled, at its asymptotic boundary, to a rigid CFT$_3$ bath. The quantum contributions to the holographic complexity of the CFT$_3$ on the brane (and in the bath) are included in the standard VC and AC prescriptions for the dCFT system; they are given in terms of classical quantities in the dual solution of a black hole on a brane in AdS$_4$. 
	
	The central feature of our construction is that it represents holographically the entire system, including the bulk quantum fields, and this allows us to incorporate the complexity of all its components. Other approaches \cite{Schneiderbauer:2019anh,Schneiderbauer:2020isp} introduce the backreaction of the quantum fields on the geometry, but not their own complexity, and this accounts for the differences in our findings about the consistency of AC.
	
	We will now recap our main conclusions.
	
	\subsubsection*{Volume Complexity}
	
	Our first result is that the VC prescription applied to this system consistently admits a semiclassical expansion of the form 
	\be\label{VC3c}
	\mathcal{C}_V(t)\, = \; \mathcal{C}^{\text{BTZ}}_V(t)\,+\, \mathcal{C}^{q}_V(t)\,+\,\dots\,,
	\ee
	where $\mathcal{C}^{q}_V \sim \hbar \,c_3$ is the leading quantum correction coming from the CFT$_3$ in the large-$c_3$ limit. The dots indicate terms which also appear in the large-$c_3$ limit, but which are suppressed by powers of the effective coupling of the theory on the brane, $g_{\text{eff}}$ (see \eqref{geff2}). Corrections to VC from higher-curvature terms in the gravitational action enter at order $g_{\text{eff}}^2$, and thus, in our study of linear order effects, we have consistently neglected them.
	
	
	
	In the introduction we presented the basic structure \eqref{genVC} of the leading corrections $\mathcal{C}^{q}_V(t)$, but now we can discern further features in it. We have identified:
	
		\begin{enumerate}
	    \item  Corrections $\delta\text{Vol}{(\Sigma)}$ to volume complexity from semiclassical backreaction on the three-dimensional geometry. These are straightforward to extract using the VC applied to the 3D quBTZ geometry.
	    \item Complexity of the quantum field state, 
	    which consists of
	    \begin{enumerate}
	        \item the complexity of the CFT$_3$ bath, $\mathcal{C}_{\text{UV}}$. It is dominated by UV physics, and for small $g_{\text{eff}}$ it is very weakly affected by the presence of a black hole in the bulk. Hence it is largely independent of the state of the dCFT, and furthermore it is time-independent, so it is of little interest to us.
	        \item the (more interesting) complexity of the CFT$_3$ on the gravitating black hole geometry. This CFT$_3$  has a cutoff from its coupling to three-dimensional gravity, and then
	        \begin{enumerate}
	            \item the CFT$_3$ degrees of freedom above the cutoff are integrated out and renormalize the gravitational constant $G_3$, which absorbs their contribution to complexity.
	            \item the CFT$_3$ degrees of freedom below the cutoff yield the proper complexity of the quantum fields $\mathcal{C}_V^{\text{bulk}}\left(\ket{\phi}\right)$.\footnote{It is difficult to separate the contribution of the bath CFT to this complexity, but we expect that it is small since for small $\ell$ the bulk black hole is far from the bath boundary. This can be seen from the fact that in this regime the stress tensor of the bath CFT is relatively small.} We have found that in our system their complexity vanishes to leading order in $g_{\text{eff}}^2$.
	        \end{enumerate}
	    \end{enumerate}
	\end{enumerate}
	
	The explicit computation of $\mathcal{C}_V^{\text{bulk}}\left(\ket{\phi}\right)$ in a thermal state of the dCFT is one of our main results. As we have argued, the fact that it vanishes is a consequence of the property that in our doubly-holographic setup, three-dimensional gravity is purely induced via integration of the UV degrees of freedom of the CFT$_3$, i.e., the bare gravitational action is zero. From the bulk viewpoint, this is the case when the brane is purely tensional. The boundary conditions that this imposes in the bulk force the VC surface to meet the brane orthogonally, and we have found that, as a result, the complexity of the quantum fields vanishes. It would be very interesting to understand this effect from the dual viewpoint of the gravity+CFT$_3$ system.
	
	From this perspective, it is natural to expect that the results will differ when there is a bare gravitational coupling. Holographically, this amounts to having an explicit Einstein-Hilbert term in the brane action, yielding an intrinsic VC term on the brane\footnote{See \cite{Chen:2020uac} for a discussion of this term for holographic entanglement entropy, or \cite{Hernandez:2020nem} for the case of subregion VC.}, which will modify the boundary conditions so that the extremal hypersurface will generically bend away from the normal. Our analysis indicates that this should result in a non-zero complexity from the quantum fields.
	
	In addition, we have checked that in the quBTZ solution the quantum-corrected VC formula \eqref{VC3c} correctly reproduces the expected computation rate for a semiclassical black hole
	\be\label{gengrowthconc}
	\left.\dfrac{d\mathcal{C}_{V}}{dt}\,\right |_{t \gg \beta} \; {\sim} \, TS_{\text{gen}}\,,
	\ee
	up to an $\mathcal{O}(1)$ coefficient that depends on the mass of the black hole.
	
	The modification of VC from semiclassical backreaction of the quantum fields is not conceptually problematic, and in principle may be computed through methods different than ours (as done in two dimensions in \cite{Schneiderbauer:2019anh}). What is much more challenging is to find a prescription for the complexity from the bulk quantum fields that applies beyond the holographic representation of these fields.
	
	
	\subsubsection*{Action Complexity}

	In contrast with this consistency of the quantum corrections to the VC proposal, assigning a proper meaning to the doubly-holographic AC of the dCFT has proven to be troublesome. The root of the problem is that the calculation of the four-dimensional action ---which is crucial for incorporating the complexity of the CFT$_3$ on the brane---does not reduce to a calculation of the three-dimensional action plus small corrections in an expansion in $g_\text{eff}$.\\ When computing the late-time rate of the AC of the system, including all orders in the effective coupling $g_\text{eff}$, we have found cancellations between terms that render the result independent of $\hbar \,c_3$ for a fixed total mass,
	\be\label{ACslopec}
	\left.\dfrac{d\mathcal{C}_A}{d{t}}\right|_{t\gg \beta}\;\sim\,\dfrac{8Mx_1\mu}{\pi}\,\; {\nsim} \, TS_{\text{gen}}\,.
	\ee
	Even worse, in the classical limit $g_\text{eff} \rightarrow 0$ the result does not reproduce the AC of the BTZ black hole. This seems troublesome, since this is the limit where the three-dimensional interpretation is most sensible. In contrast, when we evaluate the growth rate \eqref{ACslopec} in the opposite (tensionless brane) limit $g_\text{eff}\sim \ell/\ell_3 \rightarrow \infty$, the result gives the correct result for a hyperbolic Schwarzschild-AdS$_4$ black hole without a  brane -- that is, we obtain the correct behavior, but for a four-dimensional braneless system, whose dual is a defectless CFT$_3$.
	
	The observation that AC seems to effectively overlook the presence of the brane was also drawn in \cite{Chapman:2018bqj,Braccia:2019xxi} for the divergence structure of the AC in the ground state of a two dimensional dCFT/BCFT system\footnote{However, it is only in two dimensional BCFTs  that this difference has been found  \cite{Sato:2019kik}, and it is absent in other interface CFTs \cite{Auzzi:2021nrj,Auzzi:2021ozb}.}.  We have shown that the same phenomenon occurs for the late-time regime of the AC in a thermal state of a three dimensional dCFT. 
	
	However, beyond this subdued influence of the brane on AC, our analysis highlights the sensitivity of the action calculation to quantum effects near the singularity. Quantum backreaction, no matter how small, changes qualitatively, in a discontinuous manner, the nature of the inner singularity. The quantum fields near this stronger singularity then make a large contribution to AC, which does not vanish in the limit of $g_\textrm{eff}\to 0$.
	
	Qualitatively, the discrepancy between the behavior for the VC and the AC of the dCFT system comes from their markedly different geometrical character. As we take the brane close to the asymptotic boundary $g_{\text{eff}} \sim \ell/\ell_3 \ll 1$, the hyperbolic geometry makes the volume of $\Sigma_t$ to be dominated by the volume of its intersection with the brane, $\sigma_t$. This suggests that other complexity proposals that generalize the VC prescription \cite{Belin:2021bga} should behave similarly. In contrast, our results show that, first,  AC is an inherently four-dimensional object which fails to consistently furnish an effective three-dimensional AC, and second, and more importantly, AC is remarkably sensitive to discontinuous changes in the structure of the singularity. We shall not dwell on the question of whether this sensitivity is a positive or negative feature of AC. 
	
	\medskip
	
	Additionally, 
	we have analyzed the VC for a different class of states which describe quantum-dressed conical defects in AdS$_3$. We have found a quantum-corrected VC interpretation
	\be\label{VC3defc}
	\mathcal{C}_V(t)\, = \; \mathcal{C}^{\text{AdS}_3}_V\,+\, \mathcal{C}^{q}_V(t)\,+\,\dots\,,
	\ee
	where the classical part corresponds to the time-independent VC of two angular defects in AdS$_3$. The rate of VC for these objects is a completely quantum effect that is suppressed by $g_{\text{eff}}$ with respect to $TS_{\text{gen}}$. The discontinuity in the effective three-dimensional interpretation of AC that we found for BTZ black holes is even clearer in this case, since the computation rate of the conical defect fails to vanish in the classical limit.
	
	

	\subsubsection*{Other directions}

	An obvious extension of our study is to include rotation, and indeed complexity for the classical rotating BTZ black hole has been investigated in \cite{Auzzi_2018,Auzzi_2018action,Frassino:2019fgr, Bernamonti:2021jyu}.
	A detailed study of the rotating quBTZ solution was made in \cite{Emparan:2020znc}, and the four-dimensional bulk solution shares many similarities with the Kerr-AdS$_4$ black hole (actually, it is the same in the tensionless brane limit), for which AC and VC have been computed \cite{Bernamonti:2021jyu,Couch:2018phr}.
	Adding rotation might be of interest for understanding the scaling relation between complexity, entropy and the thermodynamic volume~\cite{PhysRevLett.126.101601,AlBalushi:2020ely}, but also because the character of the singularities in the interior is richer: the Cauchy horizon remains non-singular when quantum backreaction is included to leading order in $1/N$, but is expected to develop a curvature singularity with $1/N$ corrections \cite{Emparan:2020znc}. According to our findings, AC should distinctly reflect these changes.

	Another natural modification is to replace the bath, which in our study has played a subsidiary role, with another gravitating system by including a second brane in the setup. These  constructions (to our knowledge, first considered in \cite{Emparan:1999fd}) are referred to as `wedge holography' \cite{Mollabashi:2014qfa}, and represent a portion of an AdS$_D$ spacetime, with the geometry AdS$_{D-1} \times \text{(interval)}$, as dual to a CFT$_{D-2}$ (formerly the defect, now the entire quantum system). Recently they have been used in the context of the black hole information problem, e.g., in \cite{Geng:2020fxl}. Our solutions based on the AdS$_4$ C-metric can be readily adapted by placing the second brane at $r=\infty$.
	
	\section*{Acknowledgments}
	
	We would like to thank José Barbón, Alice Bernamonti, Federico Galli, César Gómez, Javier Martín-García, Juan Pedraza and Jorge Rocha for discussions. 
	We are also sincerely grateful to Le-Chen Qu for alerting us of a mistake in an earlier version of the article.
	Work for this article began while MS was visiting the Institute of Cosmos Sciences at the University of Barcelona (ICCUB), to whom he is grateful for warm hospitality during pandemic times. The work of RE, AF and MT was supported by ERC Advanced Grant GravBHs-692951, MICINN grant PID2019-105614GB-C22, AGAUR grant 2017-SGR 754, and State Research Agency of MICINN through the ``Unit of Excellence María de Maeztu 2020-2023” award to the Institute of Cosmos Sciences (CEX2019-000918-M). MS was supported by the Spanish State Research Agency (Agencia Estatal de Investigaci\'{o}n) grant to IFT Centro de Excelencia Severo Ochoa SEV-2016-0597, MINECO grant PGC2018-095976-B-C21, and FPU grant FPU16/00639. MT is also supported by the European Research Council (ERC) under the European Union’s Horizon 2020 research and innovation programme (grant agreement No 852386).
	
	
	\appendix
	
	\section{Expansion of the conformal factor}
	\label{appendix::A}
	
	In this appendix, we will show that the induced conformal factor in the VC functional $g(\ell,z) = \ell^2/(\ell +z)^3$ for $z>0$ admits the series expansion in $\ell$ given by \eqref{deltanext} in the domain $|\ell|<|z|$. To simplify notation, we will extend all functions to $z<0$ in a reflection-symmetric way along $z=0$.
	
	The leading term in the expansion of $g(\ell,z)$ is a contact term $\delta(z)$, which is responsible of localizing the VC functional onto the brane at $z=0$. As an example, consider a set of exponentially decaying test functions $f(z) = e^{-\alpha |z|}$, for which we have
	\be\label{exponentials}
	\lim\limits_{\ell\rightarrow 0}\,\int_{-\infty}^\infty \,  dz\,g(\ell,z)\,f(z)\, = \, 1\, = \,f(0)\,.
	\ee
	In fact, we could have taken any test function $f(z)$ which grows slower than $f(z) \sim z^2$ for large values of $|z|$. In this case the integral with $g(\ell,z)$ converges and we can perform the change of variables $z'= z/\ell$ to get
	\be
	\int_{-\infty}^\infty   dz\,g(\ell,z)\,f(z)\, = \,\int_{-\infty}^\infty  dz'\,g(1,z')\,f(\ell z')\, \xrightarrow[\ell\rightarrow 0]{} \, f(0)\,\int_{-\infty}^\infty  dz'\,g(1,z')\, = f(0)\,.
	\ee
	This shows that, to leading order, the conformal factor admits the expansion 
	\be
	g(\ell,z) = \delta(z) + O(\ell)\,.
	\ee
	
	The $O(\ell)$ term will also be a contact term on $z=0$. To show this, we need to evaluate the limit 
	\be
	L = \lim_{\ell\rightarrow 0}\,\int_{-\infty}^\infty dz\,\dfrac{g(\ell,z)-\delta(z)}{\ell}\,f(z)\, =\, \lim_{\ell\rightarrow 0}\,\int_{-\infty}^\infty\,dz'\,g(1,z')\,\dfrac{f(\ell z')-f(0)}{\ell}\,,
	\ee
	where in the second step we again used the variable $z' = z/\ell$ and moreover inserted the identity 
	\be
	\int_{-\infty}^\infty dz\,\dfrac{g(\ell,z)}{\ell}\,f(0) \, = \, \int_{-\infty}^\infty dz\,\dfrac{\delta(z)}{\ell}\,f(z) \,.
	\ee
	
	Defining the new test function $h(z) = (f(z)-f(0))/z$ which for large $|z|$ grows slower than $z$, we have
	\be
	L =\, \lim_{\ell\rightarrow 0}\,\int_{-\infty}^\infty dz'\,g(1,z')\,z'\,h(\ell z') \,=\, h(0)\int_{-\infty}^{\infty}dz'\,g(1,z')z'\, = h(0^+)\,=\, f'(0^+)\,,
	\ee
	which proves the result in \eqref{deltanext}. 
	
	There is an alternative proof of \eqref{deltanext} which allows to go to higher orders in $\ell$. Consider a reflection-symmetric function $f(z)$ that is differentiable on the positive real semi-axis and with the asymptotic behavior $f(z) \lesssim z$ as $z\rightarrow \infty$. Then the integral with $g(\ell,z)$ can be evaluated by parts 
	\begin{gather}
	\int_{-\infty}^\infty\,dz\,g(\ell,z)\,f(z)\, =\,2\int_{0}^\infty\,dz\,g(\ell,z)\,f(z)\, =\, \left.-\dfrac{\ell^2f(z)}{(\ell+z)^2}\right|^{\infty}_{0}\,+\,\int_{0}^\infty dz\,\dfrac{\ell^2}{(\ell+z)^2}\,f'(z)\nonumber \, = \\[.4cm] = f(0) \,+\,\ell \,\int_{0}^\infty dz\,\dfrac{\ell}{(\ell+z)^2}\,f'(z)\,.
	\end{gather}
	Performing the integration by parts again in the last integral yields
	\be
	\int_{-\infty}^\infty\,dz\,g(\ell,z)\,f(z)\, =\,f(0)\,+\,\ell\,f'(0)\,+\,\,\ell^2\int_0^\infty dz\dfrac{f''(z)}{\ell+z}\,,
	\ee
	which already proves \eqref{deltanext}. In this formalism, we can go to quadratic order by integrating by parts again
	\be
	\int_{-\infty}^\infty\,dz\,g(\ell,z)\,f(z)\, =\,f(0)\,+\,\ell\,f'(0)\,-\,\,\ell^2\log\ell\,f''(0)\,-\,\ell^2\,\int_0^\infty dz\,\log(\ell+z)\,f'''(z)\,.
	\ee
	At order $\ell^2$, we thus find two different terms. The first one is a contact term at $z=0$ which has a $\log \ell$ associated to it and the second one is the integral 
	\be
	\int_0^\infty dz\,\log(z)\,f'''(z)\,,
	\ee
	which depends on the properties of the function $f(z)$ away from $z=0$. This last regular term was expected at this order, since for the domain $|\ell| > |z|$ the function $g(\ell,z)$ has a regular power series in $\ell$ that starts at order $\ell^2$.

	\section{Late-time rate of $I(\mathcal{U}_t)$}
	\label{appendix::B}
	
	In section \ref{sec::AC}, we introduce the regularized version of AC of the system,  $\widetilde{\mathcal{C}}_A(t)$, by imposing a bulk IR-cutoff at $r=\infty$, which lies at finite proper distance from the brane. From the additivity of the action, it follows that $\mathcal{C}_A(t) = \widetilde{\mathcal{C}}_A(t)\,+\,\mathcal{C}_{\text{UV}}(t)$, where $\mathcal{C}_{\text{UV}}$ is the complexity associated to the short-range correlations of the dCFT state. Since the spacetime is static outside the black hole, it is very natural to expect that $\mathcal{C}_{\text{UV}}$ is constant at late times, and thence, all the time-dependence of AC is incorporated in $\widetilde{\mathcal{C}}_A(t)$. In this appendix, we will provide a more quantitative argument in favor of this conclusion. 
	
	As we explain in subsection \ref{sec::AC1}, we need to first introduce a second `causal diamond' $\mathcal{U}_t$ by following past/future radial null geodesics from the real asymptotic boundary at $rx = -\ell$. The WdW patch $\mathcal{W}_t$ will lie in between the two causal diamonds, $\widetilde{\mathcal{W}}_t \subset \mathcal{W}_t \subset \mathcal{U}_t$. The additivity of the renormalized action implies that: $I(\widetilde{\mathcal{W}}_t) \leq I(\mathcal{W}_t) \leq I(\mathcal{U}_t)$ \footnote{We shall think in terms of the renormalized AC obtained by adding the appropriate counterterms to the bare action. The divergent counterterms cannot depend on the state of the system, and thence cannot affect the late-time regime for the thermofield double state. Finite contributions may depend on the state, but are independent of time and will not affect the late time behavior.}.  In this section we will show that $I(\mathcal{U}_t)-I(\widetilde{\mathcal{W}}_t)$ amounts to a constant at late times. Under the reasonable assumption that $I(\mathcal{W}_t)$ is monotonic, this is enough to show that $I(\mathcal{W}_t)-I(\widetilde{\mathcal{W}}_t)$ is also constant at late times. Hence, the late-time rate of AC is given by \eqref{acasympt}, as we were anticipating from previous considerations.

	To study $I(\mathcal{U}_t)$, one needs to extend the tortoise coordinate $r_*$ in \eqref{tortoise} beyond $r=\infty$ to negative values of the $r$ coordinate
	\be
	r_*(r)\, = \, \int_{r}^{-\infty}\, \dfrac{dr}{H(r)}\,\hspace{.5cm}\text{for } r<0\,,
	\ee
	and repeat the calculation of $I(\widetilde{\mathcal{W}}_t)$ by replacing $r_m(t) \rightarrow r_m(t,x)$ for each section of fixed $x$, since now the asymptotic anchoring $r$-coordinate depends on $x$. For convenience, we will define $r_b(x) = -\ell/x$ as the asymptotic value of the $r$-coordinate for a given section of fixed $x$. The onset of computation will be the same as before, since $t_* = \min_{x}\lbrace r_*(0)-r_*(r_b)\rbrace = r_*(0)$, and $r_*(r_b) \leq 0$.
	
	\subsubsection*{Bulk contribution}
	
	In analogy with \eqref{svc}, the spacetime volume of the causal diamond $\mathcal{U}_t$ is
	\be\label{vol2}
	\text{Vol}(\mathcal{U}_t) \, = \, \text{U}_{0}\,+\,8\pi\Delta \ell^4\int_{0}^{x_1}dx\,\int_{0}^{r_m(t,x)}\,dr\,\dfrac{r^2}{(\ell + xr)^4}\,\left(t+r_*(r_b)-r_*(r)\right)\,,
	\ee
	for the constant volume
	\be
	\text{U}_{0}\,= \,16\pi\Delta \ell^4\int_{0}^{x_1}dx\,\int_{0}^{\infty}\,dr\,\dfrac{r^2}{(\ell + xr)^4}\,\left(-r_*(r_b)+r_*(r)\right)\,.
	\ee
	The large causal diamond $\mathcal{U}_t$ coincides with the regularized WdW patch $\widetilde{\mathcal{W}}_t$ on the brane, i.e. $\mathcal{U}_t \cap \textbf{brane} = \widetilde{\mathcal{W}}_t \cap \textbf{brane} = w_t$. Following the procedure of section \ref{sec::ac2}, the bulk contribution to $I(\mathcal{U}_t)$ will be the analog of \eqref{ibulk}, which is now given by
	\be\label{ibulku}
 	I_{\text{bulk}}(\mathcal{U}_t)\, = \,-\dfrac{3\,\text{Vol}({\mathcal{U}}_t)}{8\pi G_4 \ell_4^2}\,+\,\dfrac{\text{Vol}(w_t)}{4\pi G_4\ell}\,.
	\ee
	At late times, $r_m(t,x) \sim r_+$, so using \eqref{vol2} the time-derivative of the bulk contribution \eqref{ibulku} is
	\be\label{bulkutder}
	\dfrac{d}{d\bar{t}}\,I_{\text{bulk}}({\mathcal{U}}_t)\,\sim   \,-\dfrac{\Delta^2\,r_+^2\,\ell}{2G_4\ell_4^2}\,\left(1\,-\,\dfrac{\ell^2}{(\ell+x_1r_+)^2}\right)\;+\;\dfrac{\Delta^2\,r_+^2}{2G_4\ell}\,,
	\ee
	which coincides with the late-time limit of the rate of $I_{\text{bulk}}(\widetilde{\mathcal{W}}_t)$ in \eqref{bulktder}.
	
	\subsubsection*{GHY contribution}
	
	The regularized future singularity at $r = \epsilon$ will intersect each constant-$x$ section of $\mathcal{U}_t$ in a time-interval $t \in [-t_{\text{s}}(x),t_{\text{s}}(x)]$, where $t_{\text{s}}(x) = t-r_*(r_b)+r_*(\epsilon)$. The extrinsic curvature $K$ of the future singularity can be read from \eqref{extrcurv}, so the analog of \eqref{ighy} is now
	\begin{gather}
	I_{\text{GHY}}({\mathcal{U}}_t)\, = \,\lim\limits_{\epsilon\rightarrow 0} \,\dfrac{2}{8\pi G_4}\,\int_0^{2\pi \Delta}\,d\phi \int_0^{x_1}dx\,\int_{-t_{\text{s}}(x)}^{t_{\text{s}}(x)}\,dt'\,\sqrt{-H}\,\dfrac{\ell^3\,\epsilon^2}{(\ell+x\epsilon)^3}\;K\,\nonumber \\[.4cm]\, = \, \dfrac{\Delta \mu}{2\pi G_4}\,\lim\limits_{\epsilon\rightarrow 0}\int_0^{x_1}dx\,\dfrac{\ell^2}{(\ell+x\epsilon)^2}(t-r_*(r_b)+r_*(0))\,.\label{ighyu}
	\end{gather}
	The time-derivative of \eqref{ighyu} is then
	\be\label{ghyutder}
	\dfrac{dI_{\text{GHY}}({\mathcal{U}}_t)}{d\bar{t}}\,= \,\dfrac{3\Delta^2\, x_1\,\mu\,\ell }{2G_4}\,\,,
	\ee
	which coincides with the linear slope of $I_{\text{GHY}}(\widetilde{\mathcal{W}}_t)$ given by \eqref{ghytder}.
	
	\subsubsection*{Joint contribution}
	
	For the sake of brevity, we shall refer the reader to the discussion of the joint contributions in section \ref{sec::ac2} for $\widetilde{\mathcal{W}}_t$, since the discussion for $\mathcal{U}_t$ is analogous. The only joint contribution that contributes to the time-dependence comes from the lower tip of $\mathcal{U}_t$, which is now given by 
	\be
	I^{\textbf{C}}_{\text{joint}}({\mathcal{U}}_t)\, = \,-\dfrac{\Delta}{2\pi G_4}\,\int_{0}^{x_1}dx\,\dfrac{\ell^2 r_m^2}{(\ell + xr_m)^2}\,\log \left(\dfrac{\ell^2}{(\ell+xr_m)^2}\,\dfrac{|H(r_m)|}{\alpha_1\alpha_2}\right)\,,
	\ee
	where $\alpha_1,\alpha_2$ are constants which depend on the chose (affine) parametrizations of the null boundaries of $\mathcal{U}_t$. At late times, $r_m \sim r_+$, so we obtain that 
	\be\label{jointuasympt}
	\dfrac{d}{d\bar{t}}\,I_{\text{joint}} ({\mathcal{U}}_t)\, \sim \,\dfrac{\Delta^2 \ell x_1r_+^2H'(r_+) }{2 G_4 (\ell+x_1r_+) }\, = \, 2\,TS_{\text{gen}}\,,
	\ee
	in complete analogy with \eqref{jointasympt}.
	
	We have thus shown that all the relevant terms in $I(\mathcal{U}_t)$ have the same late-time slope to the ones in $I(\widetilde{\mathcal{W}}_t)$. The late-time regime of $I(\mathcal{U}_t)-I(\widetilde{\mathcal{W}}_t)$ is thus a constant, which from the previous consideration means that the value of AC has to grow with the slope \eqref{acasympt}.

\bibliographystyle{style}
\bibliography{CQBTZ.bib}

\end{document}